\documentclass[11pt,a4paper]{article}
\pdfoutput=1

\usepackage{tikz}
\usetikzlibrary{shapes,arrows}
\usepackage{jcappub}
\usepackage[utf8]{inputenc}
\usepackage{graphicx}
\usepackage{color}
\usepackage{amsmath}
\usepackage[normalem]{ulem}
\usepackage{todonotes}
\usepackage{menukeys}

\newcommand{\be}{\begin{equation}} 
\newcommand{\ee}{\end{equation}}
\newcommand{\bea}{\begin{eqnarray}} 
\newcommand{\eea}{\end{eqnarray}}
\newcommand{\td}{{\rm d}}
\newcommand{\higgs}{\mathcal{H}}

\title{Hidden Inflaton Dark Matter}

\author[a]{Juan P. Beltr\'{a}n Almeida,}
\author[b]{Nicol\'{a}s Bernal,}
\author[c,d]{Javier Rubio}
\author[e,f]{and Tommi Tenkanen}

\affiliation[a]{Departamento de F\'isica, Universidad Antonio Nari\~{n}o,\\
Carrera 3 Este \# 47A-15, Bogot\'{a}, Colombia}
\affiliation[b]{Centro de Investigaciones, Universidad Antonio Nari\~{n}o,\\
Carrera 3 Este \# 47A-15, Bogot\'{a}, Colombia}
\affiliation[c]{Institut f\"ur Theoretische Physik, Ruprecht-Karls-Universit\"at Heidelberg,\\
Philosophenweg 16, 69120 Heidelberg, Germany}
\affiliation[d]{Department of Physics and Helsinki Institute of Physics, \\ PL 64, FI-00014 University of Helsinki, Finland}
\affiliation[e]{Department of Physics and Astronomy, Johns Hopkins University, \\
Baltimore, MD 21218, United States of America}
\affiliation[f]{Astronomy Unit, Queen Mary University of London,
 \\ Mile End Road, London, E1 4NS, United Kingdom}

\emailAdd{juanpbeltran@uan.edu.co}   
\emailAdd{nicolas.bernal@uan.edu.co}
\emailAdd{javier.rubio@helsinki.fi}
\emailAdd{ttenkan1@jhu.edu}

\abstract{If cosmic inflation was driven by an electrically neutral scalar field stable on cosmological time scales, the field necessarily constitutes all or part of dark matter (DM). We study this possibility in a scenario where the inflaton field $s$ resides in a hidden sector, which is coupled to the Standard Model sector through the Higgs portal $\lambda_{hs} s^2\higgs^\dagger\higgs$ and non-minimally to gravity via $\xi_s s^2 R$. We study scenarios where the field $s$ first drives inflation, then reheats the Universe, and later constitutes all DM. We consider two benchmark scenarios where the DM abundance is generated either by production during reheating or via non-thermal freeze-in. In both cases, we take into account all production channels relevant for DM in the mass range from keV to PeV scale. On the inflationary side, we compare the dynamics and the relevant observables in two different but well-motivated theories of gravity (metric and Palatini), discuss multifield effects in case both fields ($s$ and $h$) were dynamical during inflation, and take into account the non-perturbative nature of particle production during reheating. We find that, depending on the initial conditions for inflation, couplings and the DM mass, the scenario works well especially for large DM masses, $10^2$~GeV$\lesssim m_{s}\lesssim 10^6$~GeV, although there are also small observationally allowed windows at the keV and MeV  scales. We discuss how the model can be tested through astrophysical observations.
}

\begin{document}

\begin{flushright}
PI/UAN-2018-641FT \\
HIP-2018-29/TH
\end{flushright}

\maketitle

\section{Introduction}

If cosmic inflation was driven by an electrically neutral inflaton scalar field stable on cosmological scales, this field necessarily constitutes a dark matter (DM) component~\cite{Liddle:2006qz,Cardenas:2007xh,Panotopoulos:2007ri,Liddle:2008bm,Bose,Lerner:2009xg,DeSantiago:2011qb,Khoze:2013uia,Mukaida:2014kpa,Fairbairn:2014zta,Bastero-Gil:2015lga,Kahlhoefer:2015jma,Tenkanen:2016twd,Daido:2017wwb,Choubey:2017hsq,Daido:2017tbr,Hooper:2018buz,Borah:2018rca,Manso:2018cba,Rosa:2018iff}. This can be the case either because the reheating stage after inflation was not complete and left behind a remnant of the scalar condensate, or because the inflaton excitations were eventually created by the Standard Model (SM) products following the decay of the inflaton zero mode. The scenario is particularly appealing since it is able to explain two things at once: the origin of the DM component and the generation of the primordial curvature perturbations leading to temperature fluctuations in the Cosmic Microwave Background (CMB)~\cite{Aghanim:2018eyx}.

In this paper we consider models where decay of the homogeneous inflaton condensate after inflation is always complete. The inflaton {\it particles} can still be stable, as reheating is not, in general, a process where individual particles decay, transferring their energy into other particles one by one, but rather a non-perturbative process where a time-dependent condensate transfers its energy density into other fields (see e.g. Refs.~\cite{Kofman:1994rk,Kofman:1997yn,Mukaida:2013xxa}). If the coupling between the inflaton and the SM sector is large enough, the inflaton particles will enter in thermal equilibrium with the other particles produced in reheating. Then, at some point when their mutual interaction rate cannot keep up anymore with the expansion of the Universe, the inflaton particles undergo thermal freeze-out. If the inflaton particles were stable, they will constitute the usual Weakly Interacting Massive Particle (WIMP) DM. On the other hand, if the coupling between the inflaton and the SM sector was very small, the inflaton field may have reheated the Universe without ever entering into thermal equilibrium with the resulting heat bath itself. In this case, however, the inflaton particles may still have been produced by the freeze-in mechanism~\cite{McDonald:2001vt,Hall:2009bx} after inflation and constitute all the DM component. In this mechanism, DM particles are produced non-thermally by decays and annihilations of SM particles, without subsequent thermalization of them with the SM sector. As this necessarily requires a very small coupling between DM and the SM sector, the corresponding DM particle is often dubbed Feebly Interacting Massive Particle (FIMP).

Given the increasingly stringent observational constraints on WIMP DM~\cite{Arcadi:2017kky}, the freeze-in mechanism has recently become more and more popular as a production mechanism for DM; for a recent review of FIMP models and constraints, see Ref.~\cite{Bernal:2017kxu} (see also~Ref.~\cite{Belanger:2018mqt} for the implementation of this mechanism into the micrOMEGAs code). In this paper, we will study exactly this scenario. The model we will study is arguably one of the most economic models for explaining both DM and inflation, as all we need is a real $\mathbb{Z}_2$-symmetric singlet scalar feebly interacting with the SM sector via the Higgs portal coupling.\footnote{In our case, we simply assume this to be the case, as it is phenomenologically interesting. Further motivation for small couplings can be found in e.g. higher symmetry groups~\cite{Alanne:2016mmn,Alanne:2016mpa}, clockwork mechanism~\cite{Kim:2017mtc,Goudelis:2018xqi}, or in the requirement of preserving flatness of the inflationary potential~\cite{Enqvist:2016mqj,Rusak:2018kel}.} On top of this coupling, we will assume that the scalar is only weakly self-interacting and has a non-minimal coupling to gravity.

The non-minimal coupling to gravity is not only allowed by the model symmetries but also generated radiatively even if it was initially set to zero~\cite{Birrell:1982ix}. This is exactly what models like Higgs inflation~\cite{Bezrukov:2007ep} utilize very successfully, as such models are in very good agreement with the most recent observations of the CMB~\cite{Akrami:2018odb} (for a review, see Ref.~\cite{Rubio:2018ogq}). However, in contrast to other works on unification of inflation and DM, in this paper we will study our scenario in both {\it metric} and {\it Palatini} counterparts of gravity. As recently studied in a number of works, this choice plays an important role in determining the field dynamics during inflation and the resulting predictions for observables, such as the ratio of tensor to scalar perturbations~\cite{Bauer:2008zj,Tamanini:2010uq,Rasanen:2017ivk,Fu:2017iqg,Tenkanen:2017jih,Racioppi:2017spw,Markkanen:2017tun,Jarv:2017azx,Racioppi:2018zoy,Enckell:2018kkc,Carrilho:2018ffi,Enckell:2018hmo,Antoniadis:2018ywb,Rasanen:2018fom,Kannike:2018zwn,Rasanen:2018ihz}. 
What separates these two scenarios is the choice of gravitational degrees of freedom. In the metric counterpart of gravity, one {\it assumes} that the space-time connection is given uniquely by the metric only, i.e. that the connection is the usual Levi-Civita one, whereas in the Palatini counterpart one allows the connection to be an a priori free parameter, whose constraint equation determines how it depends on the metric and the matter content of the theory. Notably, in General Relativity (GR) these two approaches render to mere reformulations of the same theory. However, when non-minimal couplings between matter fields and gravity are explicit, these two approaches describe two inherently different theories of gravity (for extended discussion on the topic, see e.g. Ref.~\cite{Sotiriou:2008rp}). As in inflationary scenarios where the scalar field which is coupled non-minimally to gravity relaxes down to very small values after inflation, in the present Universe there are no ways of distinguishing between these two theories (given that the gravity sector remains otherwise unchanged from that of GR). However, as shown in the above works, this choice of the underlying theory of gravity can have an important effect on the dynamics during inflation. In this paper, through the unification of inflation and DM, we show that this choice can also affect the physics after inflation.

This paper is organized as follows. In Section~\ref{model}, we introduce the model and discuss the different scenarios considered in this paper. The details of the inflationary stage are presented in Section~\ref{inflation}, where we consider 
both metric and Palatini theories and perform an analysis of the inflationary dynamics. Section~\ref{reheating} is devoted to the study of the  reheating stage in different scenarios. The DM production, both via freeze-in and reheating, is discussed in Section~\ref{DMproduction}. Finally, our conclusions are presented in Section~\ref{conclusions}.

\section{The model}
\label{model}

We study a minimalistic extension of the SM of particle physics where, on top of the SM particle content, we include a real scalar singlet $s$, playing the double role of the inflaton field and the DM component. In order for this new degree of freedom to be a viable DM candidate, we require the theory to exhibit a discrete $\mathbb{Z}_2$ symmetry.\footnote{For a recent study where the DM scalar $s$ is not absolutely stable, see Ref.~\cite{Heeba:2018wtf}. Our findings can be easily generalized to account for this possibility, cf.~Section~\ref{DMproduction}.} 
In particular, we consider a scalar potential 
\be
\label{potential}
V(\higgs,s) = \mu_{h}^2\higgs^\dagger\higgs+\lambda_{h}(\higgs^\dagger\higgs)^2+\frac{\mu_{s}^2}{2} s^2+\frac{\lambda_{s}}{4}s^4+\frac{\lambda_{hs}}{2}\higgs^\dagger\higgs s^2 \,,
\ee
with $\sqrt{2}\higgs^{\rm T}=(0,\,h)$ the SM $SU(2)$ gauge doublet in the unitary gauge, $v=246$~GeV the vacuum expectation value of the SM Higgs field and  $\lambda_{h}$ and $\lambda_{s}$  two positive-definite couplings ensuring vacuum stability. Additionally, we allow both $\higgs$ and $s$ to interact non-minimally with gravity by including a term
\be\label{Vgravity}
\delta S = \frac{1}{2}\int d^4 x \sqrt{-g}\left[\left( \xi_{s}s^2 + 2 \,\xi_{h}\higgs^\dagger\higgs\right) g^{\mu\nu}R_{\mu\nu}(\Gamma) \right] 
\ee
on top of the usual Einstein-Hilbert action, with $g_{\mu\nu}$ the metric tensor and $\Gamma$ an arbitrary connection. In the so-called metric theory, the connection $\Gamma$ is identified with the Levi-Civita connection
\be\label{LCconnection}
	\bar{\Gamma}^\lambda_{\alpha\beta} = \frac{1}{2}g^{\lambda\rho}(\partial_\alpha g_{\beta\rho} + \partial_\beta g_{\rho\alpha} - \partial_\rho g_{\alpha\beta}) \,.
\ee
In the alternative Palatini approach, the metric and the connection are rather treated as independent variables.
However, here we assume, for simplicity, that the connection is torsion-free $\Gamma^\lambda_{\alpha\beta}=\Gamma^\lambda_{\beta\alpha}$ (for non-vanishing torsion, see Ref. \cite{Rasanen:2018ihz}).

For a Friedmann-Lema\^itre-Robertson-Walker cosmology, Eq.~\eqref{Vgravity} can be interpreted as a Hubble-induced mass term for the $s$ and $\higgs$ fields. While this gravitational contribution plays a very important role during inflation, it can be safely neglected in the late Universe, where the curvature is very small. At low energies, the effective square mass of the scalar field $s$ can be well-approximated by the sum of the mass parameter $\mu_s^2$ and the \textit{Higgs portal} contribution $\lambda_{hs}\higgs^\dagger\higgs$.  In order to ensure that the $\mathbb{Z}_2$ symmetry is not spontaneously broken at the electroweak vacuum $v$, we demand this mass to be positive definite, 
\begin{equation}
m_{s}^2\equiv  \mu_{s}^2 + \lambda_{hs}\,v^2/2>0\,.
\end{equation}
    \tikzstyle{block2} = [ellipse,fill=red!20, draw, 
    text width=8em, text centered, rounded corners, minimum height=2em,node distance=1cm]
\tikzstyle{block1} = [rectangle, draw, fill=blue!20, 
    text width=7em, text centered, rounded corners, minimum height=2em,node distance=5cm]
  \tikzstyle{decision} = [diamond, draw, fill=blue!20, 
    text width=3em, text badly centered, node distance=2cm, inner sep=0pt]
\tikzstyle{block} = [rectangle, draw, fill=green!20, 
    text width=6em, text centered, rounded corners, minimum height=2em,node distance=1.7cm]
\tikzstyle{line} = [draw, -latex']
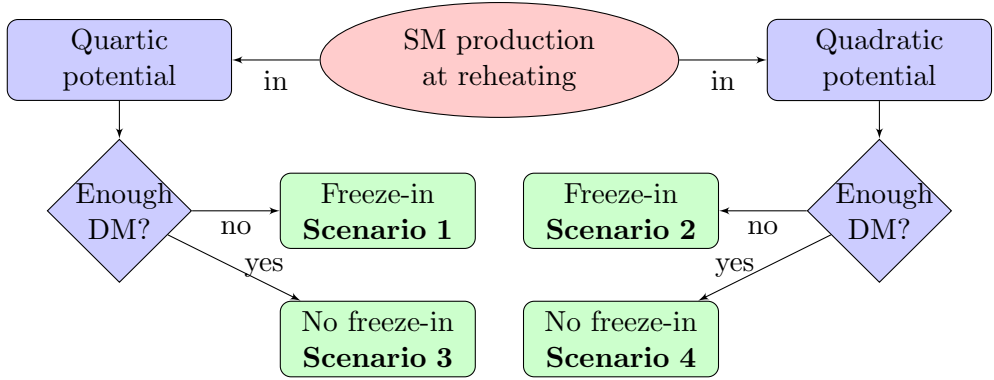
\begin{figure}
\centering
\begin{tikzpicture}[node distance = 2cm, auto]
    \node [block2] (init) {SM production at reheating};
    \node [block1, left of=init] (quartic) {Quartic potential};
    \node [block1, right of=init] (quadratic) {Quadratic potential}; 
\node [decision, below of=quartic] (enoughQ) {Enough DM?};
\node [block, below of=enoughQ, right of=enoughQ, right of=enoughQ] (freezeinN) {No freeze-in {\bf Scenario 3}};
\node [block, right of=enoughQ,right of=enoughQ] (freezeinY) {Freeze-in {\bf Scenario 1}};
\node [decision, below of=quadratic] (enoughQ2) {Enough DM?};
\node [block, left of=enoughQ2,left of=enoughQ2,below of=enoughQ2] (freezeinN2) {No freeze-in {\bf Scenario 4}};
\node [block, left of=enoughQ2,left of=enoughQ2] (freezeinY2) {Freeze-in {\bf Scenario 2}};
\path [line] (init) -- node[below] {in}(quadratic);
\path [line] (init) -- node[below] {in} (quartic);
    \path [line] (quartic) -- (enoughQ);
     \path [line] (enoughQ) -- node[right] {yes} (freezeinN);
    \path [line] (enoughQ) -- node[below] {no} (freezeinY);
 \path [line] (quadratic) -- (enoughQ2);
       \path [line] (enoughQ2)  -- node[below] {no} (freezeinY2);
    \path [line] (enoughQ2) -- node[left] {yes} (freezeinN2);
\end{tikzpicture}
\caption{The four non-thermal DM production mechanisms considered in this paper.}\label{fig:flow}
\end{figure}
We will study four phenomenologically distinct scenarios in which the DM component is non-thermally produced (cf. Fig.~\ref{fig:flow}):
\begin{enumerate}
\item Reheating after inflation occurs in an effectively quartic potential,  $V\simeq \lambda_{s}s^4/4$ and no significant amount of DM is produced during it. All DM is produced by freeze-in after reheating.

\item Reheating occurs only when the inflaton field has relaxed to the quadratic part of its potential, $V\simeq m_{s}^2s^2/2$, and no significant amount of DM is produced during it. All DM is produced by freeze-in after reheating.

\item Reheating occurs in an effectively quartic potential and while all the observed DM is produced during it, the inflaton decays dominantly into the SM sector. No significant amount of DM is produced by freeze-in after reheating.

\item Reheating occurs only when the inflaton field has relaxed to the quadratic part of its potential. Again, all the observed DM is produced during it but the inflaton decays dominantly into the SM sector. No significant amount of DM is produced by freeze-in after reheating.
\end{enumerate}
In all cases, we assume that the inflaton field interacts so feebly with the SM sector that it never enters into thermal equilibrium with it. The scenario $1$ has previously been studied in Ref.~\cite{Tenkanen:2016twd}, whereas scenarios $2$ to $4$ are new to the best of our knowledge. To set up the stage for DM production during reheating and/or freeze-in, we begin by studying the dynamics during inflation.

\section{Cosmic inflation}
\label{inflation}

In this paper we are mainly interested in a scenario in which the scalar potential~\eqref{potential} is dominated by the $\lambda_s s^4/4$ term. There are two ways  in which this can happen: either inflation starts around the $h=0$ direction or the couplings are sufficiently hierarchical at the inflationary scale, namely  $\lambda_h\ll \lambda_s\simeq 10^{-8}$ (cf. Appendix~\ref{multifield_appendix}). 
While the latest possibility could be achieved  with a suitable running of the couplings (see for instance Ref.~\cite{Degrassi:2012ry}), it would require considerable fine-tuning of the SM properties at the electroweak scale. In the main text we will therefore assume that $h=0$ initially,  such that inflation occurs in the $s$-direction. As shown below, this corresponds to the usual limit in which the amplitude of the Einstein-frame potential at large field values is mainly controlled by the combination $\lambda_s/\xi^2_s$~\cite{Liddle:2006qz,Cardenas:2007xh,Panotopoulos:2007ri,Liddle:2008bm,Lerner:2009xg,Bose,DeSantiago:2011qb,Khoze:2013uia,Bastero-Gil:2015lga,Kahlhoefer:2015jma,Tenkanen:2016twd,Choubey:2017hsq}.\footnote{Other choices would rather correspond to Higgs inflation~\cite{Rubio:2018ogq} or to Higgs-portal driven inflation. For an extensive discussion of the different possibilities the reader is referred to Refs.~\cite{Lebedev:2011aq,Kaiser:2013sna,Schutz:2013fua,Ballesteros:2016xej}.} We will justify the choice $h=0$ \textit{a posteriori} by showing that this is indeed the phenomenologically interesting limit where the same field can both drive inflation and, at a later stage, provide a FIMP DM candidate. 

We emphasize that for general expectation values $h\neq 0$, the inflationary stage should be treated as a multifield scenario and will consequently extend the existing formalism to account for the non-vanishing value of the Higgs and its non-minimal coupling to gravity in Appendix~\ref{multifield_appendix}. Note,  however,  that for the particular case under consideration $h\to 0$, meaning that the Higgs field becomes an energetically subdominant spectator field with Planck suppressed fluctuations $(H_*/M_P)^2$~\cite{Enqvist:2013kaa,Espinosa:2015qea}. This allows us to approximate the dynamics by that of a single-field scenario with action 
\be \label{nonminimal_action1}
	S_J = \int d^4x \sqrt{-g}\left(\frac{1}{2}\left(M_{P}^2 + \xi_s s^2\right) g^{\mu\nu}R_{\mu\nu}(\Gamma) + \frac{1}{2} g^{\mu\nu}\partial_{\mu}s\partial_{\nu}s - V(s) \right) \,.
\ee
and potential $V(s)\simeq \lambda_s/4\,s^4$.
The non-minimal coupling to gravity in this expression can be removed by performing a Weyl transformation
\be \label{Omega}
	g_{\mu\nu} \to \Omega(s)^{2}g_{\mu\nu}, \hspace{2cm} \Omega^2(s)\equiv 1+\frac{\xi_s s^2}{M_{P}^2} \,,
\ee
which gives the Einstein-frame action
\be \label{einsteinframe1}
	S_E = \int d^4x \sqrt{-g}\left(\frac{M_{P}^2}{2}g^{\mu\nu}R_{\mu\nu}(\Gamma_E)-\frac{1}{2}\frac{M_{P}^2+\xi_s s^2+
	6\alpha \xi_s^2s^2}{\Omega^2(s)\left(M_{P}^2+\xi_s s^2\right)}\, g^{\mu\nu} \partial_{\mu}s\partial_{\nu}s - \frac{V(s)}{\Omega(s)^4} \right) \,,
\ee
with  $\alpha=1$ in the metric case and $\alpha=0$ in the Palatini case. Note that in this frame the gravitational part of the action takes the usual Einstein-Hilbert form, allowing to identify the Einstein-frame connection $\Gamma_E$ with the Levi-Civita connection $\bar \Gamma$ in Eq.~\eqref{LCconnection}. With a suitable field redefinition 
\be \label{chi1}
	\frac{ds}{d\chi} = \sqrt{\frac{\Omega^2(s)\left(M_{P}^2+\xi_s s^2\right)}{M_{P}^2+\xi_s s^2+6\alpha \xi_s^2s^2}} \,,
	\ee
the kinetic term in Eq.~\eqref{einsteinframe1} can be expressed in a canonical form. The solution of this differential equation takes the form \cite{GarciaBellido:2008ab,Bauer:2008zj,Rasanen:2017ivk}
\be
\label{chi_solution}
\frac{\sqrt{\xi_s}}{M_{P}}\chi = \sqrt{1+6\alpha\xi_s}\sinh^{-1}\left(\sqrt{1+6\alpha\xi_s}u\right) - \sqrt{6\xi}\alpha\sinh^{-1}\left(\sqrt{6\xi_s}\frac{u}{\sqrt{1+u^2}}\right) ,
\ee
with $u\equiv \sqrt{\xi_s}s/M_{P}$. In terms of the new $\chi$ variable, the action~\eqref{einsteinframe1} reads
\be \label{EframeS1}
	S_{\rm E} = \int d^4x \sqrt{-g}\bigg(\frac{1}{2}M_{P}^2R -\frac{1}{2}{\partial}_{\mu}\chi{\partial}^{\mu}\chi - U(\chi)  \bigg) \,,
\ee
with $U(\chi) =V(s(\chi))/ \Omega^{4}(s(\chi))$ and $R = g^{\mu\nu}R_{\mu\nu}(\bar{\Gamma})$. Note that that when $\chi\to 0$ (corresponding to $s\to 0$ or $\Omega\to 1$), the usual Einstein-Hilbert term of GR is recovered, regardless of the choice of formalism (metric or Palatini). The canonically normalized field can be expressed as
\bea
s(\chi) 
\begin{cases} 
\simeq \displaystyle\frac{M_{P}}{\sqrt{\xi_s}} \exp\left(\sqrt{\frac{1}{6}}\frac{\chi}{M_{P}} \right)	  & \quad \mathrm{metric} ,\\   
= \displaystyle\frac{M_{P}}{\sqrt{\xi_s}}\sinh\left(\frac{\sqrt{\xi_s}\chi}{M_{P}}\right)& \quad \mathrm{Palatini} , 
\end{cases}
\eea
and hence the large field Einstein frame potential becomes
\bea \label{chipotential1}
	U(\chi) =\frac{\lambda_s}{4} \frac{s^4(\chi)}{\Omega^4(s(\chi))} 
\begin{cases}	
\simeq \displaystyle\frac{\lambda_s M_{P}^4}{4\xi_s^2}
\left[1+\exp\left(-\sqrt{\frac{2}{3}} \displaystyle\frac{\chi}{M_{P}} \right) \right]^{-2}
	& \quad \mathrm{metric} ,\\ 
	= \displaystyle\frac{\lambda_s M_{P}^4}{4\xi_s^2}
	\tanh^4\left(\displaystyle\frac{\sqrt{\xi_s}\chi}{M_{P}}\right)& \quad \mathrm{Palatini}. 
\end{cases}
\eea
Note that the expressions in the metric case apply for $\xi_s\gg 1$ and $\chi\gg M_{P}/\xi_s$, whereas the expressions in the Palatini case are exact. However, in our numerical analysis we do not use the approximate result \eqref{chipotential1} but rather compute everything using the exact result $U(\chi) =V(s(\chi))/ \Omega^{4}(s(\chi))$ with $\Omega(s)$ given by Eq.~\eqref{Omega} and $s(\chi)$ by Eq.~\eqref{chi1}. 

The amplitude of primordial spectrum of curvature perturbations is given by~\cite{Lyth:1998xn,Aghanim:2018eyx}
\be
\label{cobe}
A_s = \frac{1}{24 \pi^2 M_{P}^4} \frac{U(\chi_i)}{\epsilon(\chi_i)} ,
\ee
which relates the non-minimal coupling to the number of required $e$-folds and the quartic self-interaction as
\be \label{xicondition1}
	\xi_s \simeq 
	\begin{cases}
		\sqrt{\displaystyle\frac{\lambda_s}{72\pi^2 A_s}}N & \quad \mathrm{metric} , \vspace{3mm} \\
		\displaystyle\frac{\lambda_s N^2}{12\pi^2 A_s} & \quad \mathrm{Palatini} . \\
	\end{cases}
\ee
The observed amplitude is $\mathcal{P}_{\zeta}=2.1\times 10^{-9}$ \cite{Akrami:2018odb}. The inflationary dynamics is also characterized by the spectral tilt of the primordial spectrum of curvature perturbations, its running and the tensor-to-scalar ratio, namely 
\be
\label{nsralpha}
n_s \simeq 1 - 6\epsilon + 2\eta\,,\quad
\alpha_s \simeq -24\epsilon^2 + 16\epsilon\eta -2\delta\,,\quad
r\simeq 16\epsilon\,,
\ee
with 
\be
\label{SRparameters1}
	\epsilon \equiv \frac{1}{2}M_{ P}^2 \left(\frac{U'}{U}\right)^2 \,, \hspace{10mm}
	\eta \equiv M_{P}^2 \frac{U''}{U} \,, \hspace{10mm}
	\delta \equiv M_{\rm P}^4 \frac{U'}{U}\frac{U'''}{U} \,, 
\ee
the usual slow-roll parameters and the primes denoting derivatives with respect to $\chi$. These quantities are assumed to be evaluated at the number of $e$-folds \be \label{Ndef}
	N_* = \frac{1}{M_{P}^2} \int_{\chi_f}^{\chi_i} {\rm d}\chi \, U \left(\frac{{\rm d}U}{{\rm d} \chi}\right)^{-1},
\ee
at which the pivot scale crosses the horizon, with $\chi_f$ the field value at the end of inflation, $\epsilon(\chi_f)\equiv 1$. In the limit of large $N_*$, we obtain
\be \label{nsr2app}
r\simeq \frac{2}{\vert \kappa_c\vert N_*^2}\,, \hspace{15mm}
n_s \simeq 1-\frac{2}{N_*} \,,  \hspace{15mm} 
\alpha_s \simeq  - \frac{2}{N_*^2}\,,
\ee 
where $\kappa_c \equiv-\xi_{s}/(1+6\,\alpha\,\xi_{s})$. 

\begin{figure}[t]
\begin{center} 
\includegraphics[height=0.33\textwidth]{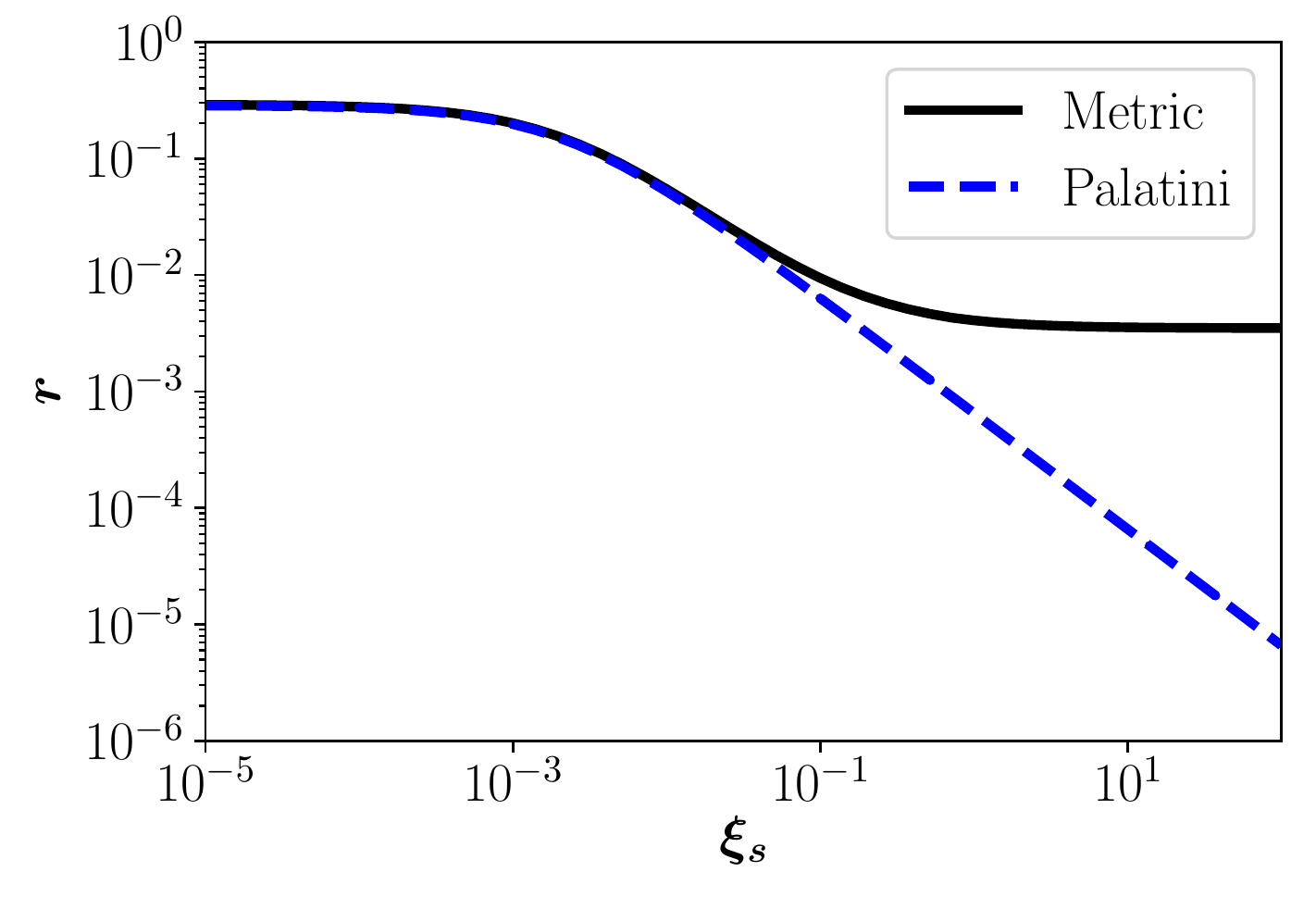} 
\includegraphics[height=0.33\textwidth]{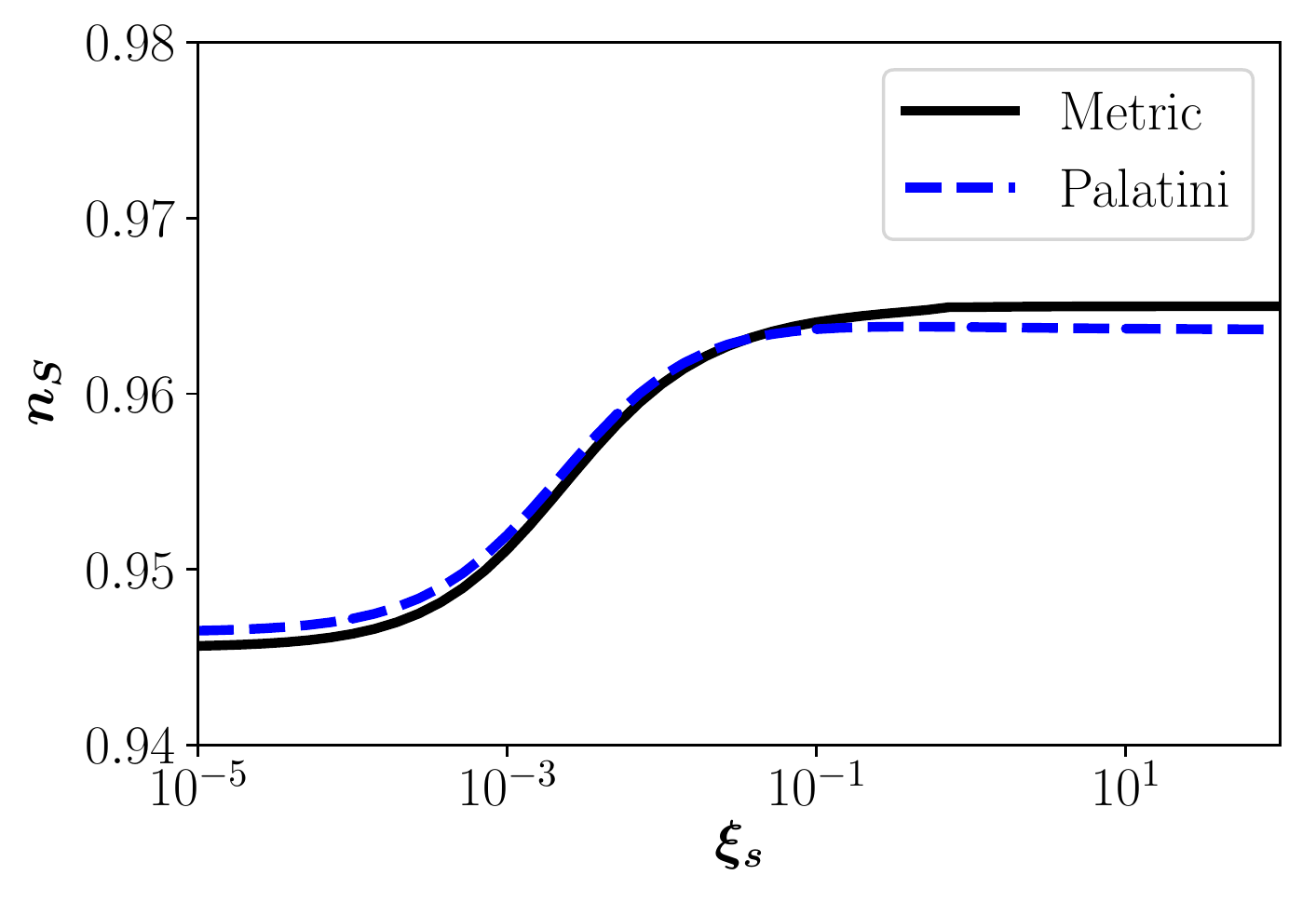}
\includegraphics[height=0.33\textwidth]{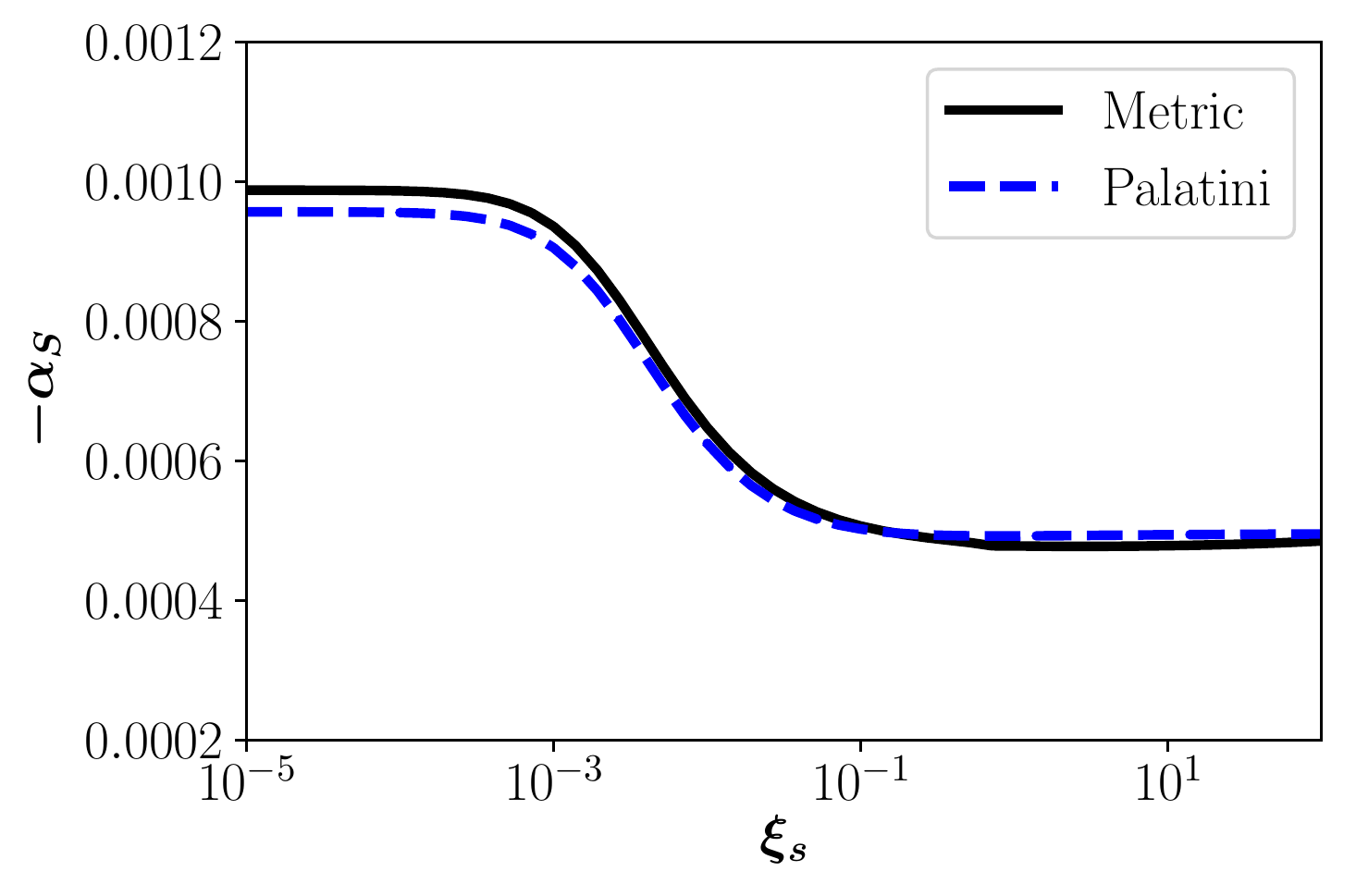} 
\includegraphics[height=0.33\textwidth]{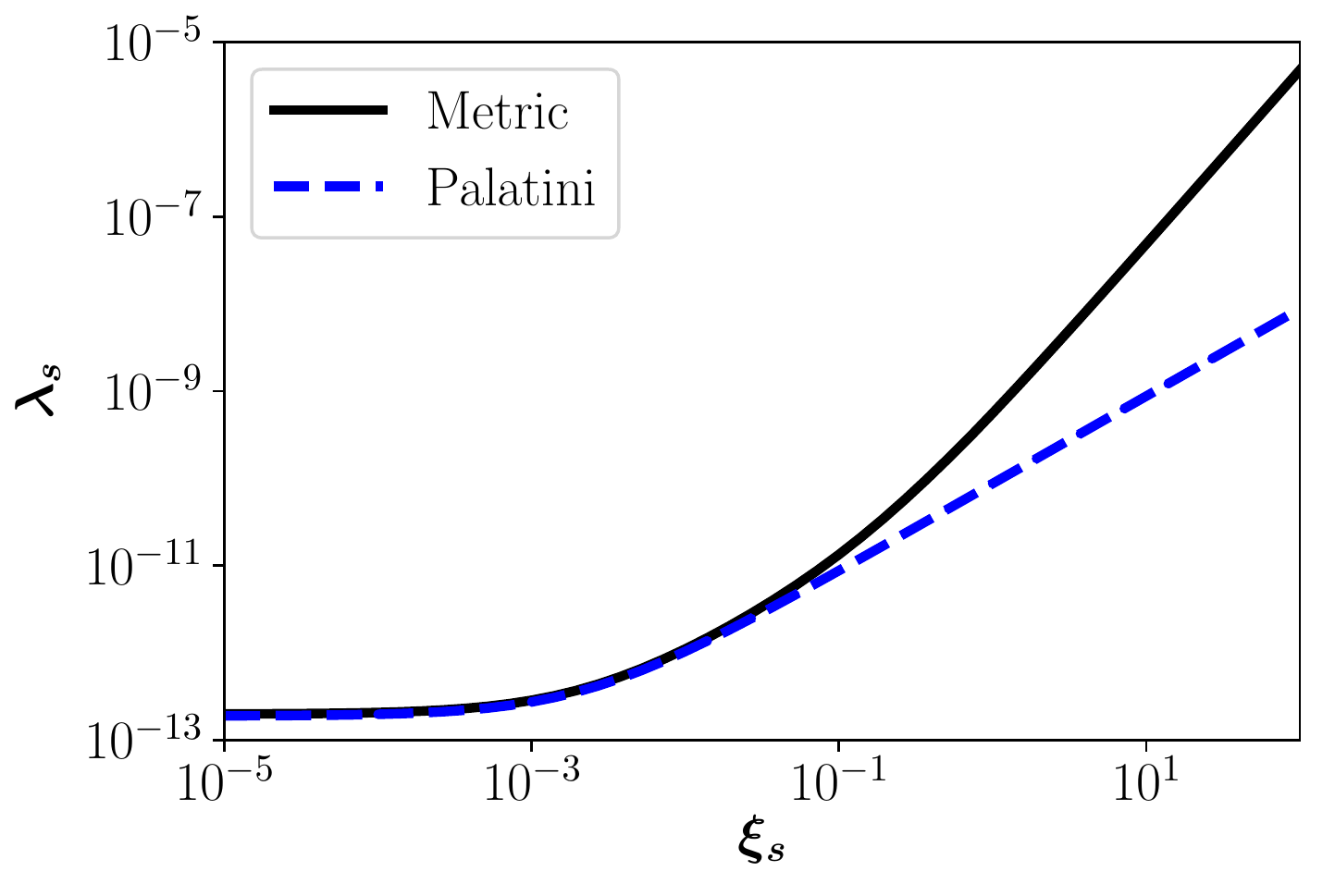} 
\includegraphics[height=0.33\textwidth]{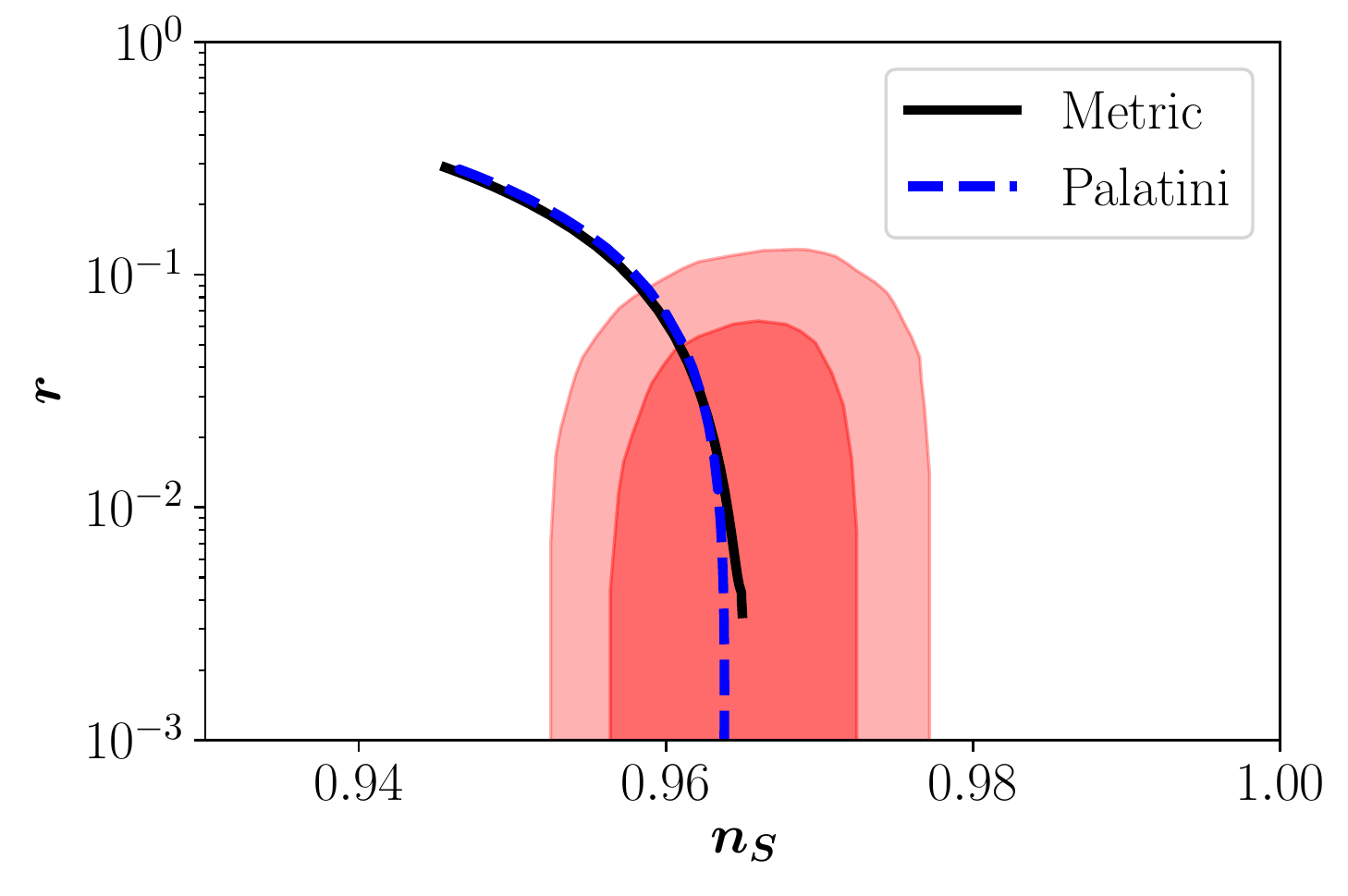}
\caption{Upper four panels: The dependence of the tensor-to-scalar ratio $r$, the spectral index $n_s$, its running $\alpha_s$ and the self-coupling $\lambda_s$ on the non-minimal coupling to gravity $\xi_{s}$, for $N_*=55$ $e$-folds. The lower panel shows the variation of the model prediction in the $n_s-r$ plane for different values of $\xi_{s}$.
The red bands correspond to the $68\%$ (dark) and $95\%$~CL (light) regions from Planck. The black and blue lines show the metric and Palatini theories, respectively.
}
 \label{slow-roll}
\end{center}
\end{figure}

The dependence of the tensor-to-scalar ratio, the spectral index, its running, and the self-coupling $\lambda_{s}$ on the non-minimal coupling to gravity $\xi_{s}$ is depicted in Fig.~\ref{slow-roll}, where for illustration purposes we consider a fiducial value $N_*=55$.
The black and blue lines correspond to the metric and Palatini theories, respectively. Additionally, the bottom panel shows the variation of the model predictions in the $n_s-r$ plane for different values of $\xi_{s}$.
The red bands correspond to the Planck 2018 $68\%$ (dark) and $95\%$~CL (light) regions~\cite{Akrami:2018odb}. In the metric case the predicted tensor-to-scalar ratio is always $r \gtrsim 10^{-3}$ and hence well within the reach of current or future planned experiments such as BICEP3~\cite{Wu:2016hul}, LiteBIRD~\cite{Matsumura:2013aja} and the Simons Observatory~\cite{Ade:2018sbj}. In the Palatini case, however, the predicted tensor-to-scalar ratio is within reach of current or future experiments only if $\xi_{s}\lesssim10$, which requires a relatively small self-coupling $\lambda_{s} \lesssim 10^{-9}$. However, as we will see in the following sections, this is exactly what it is {\it required} when the inflaton couples only very weakly to the SM sector and acts as a FIMP DM candidate. 

Note that all the above considerations are based on a tree-level analysis. The computation of radiative corrections in non-minimally coupled theories is a subtle issue due to their intrinsic non-renormalizability (see Ref.~\cite{Rubio:2018ogq} for a review). In particular, any sensible computation within this framework requires the inclusion of an infinite number of higher-dimensional operators, which can be either associated to new physics or generated by the theory itself via radiative corrections~\cite{Bezrukov:2012hx}.  We postpone the inclusion of quantum effects, both in the metric and Palatini scenarios, to a future publication. 
\section{Reheating after inflation}
\label{reheating} 

As discussed above, the predictions for the inflationary observables in DM-driven inflation depend on the number of $e$-folds between horizon exit and the end of inflation. This quantity, in turn, depends on the  whole post-inflationary history and, in particular, on the duration of the reheating stage immediately following the end of inflation. Regarding this period, there are two necessary requirements for the $s$ particles to play simultaneously the role of the inflaton and the DM component,  namely $i)$ the energy transfer must occur dominantly through the SM sector and $ii)$  the produced particles must attain thermal equilibrium at a temperature $T_{\rm{RH}}$ much larger than the temperature $T_{\rm{DM}}$ at which DM is produced. 

In the following, we will study the reheating stage in both the metric and Palatini theories.

\subsection{The metric case}

The entropy production following the end of $s$-inflation in its metric formulation has been studied exhaustively in the literature~\cite{Lerner:2011ge,Tenkanen:2016idg,Tenkanen:2016twd} (see also Refs.~\cite{Bezrukov:2008ut,GarciaBellido:2008ab}). Here we will simply review the main results. 

Soon after the end of inflation, the inflaton field begins to oscillate around the minimum of its effective potential,  which smoothly interpolates between a quadratic potential \cite{Bezrukov:2008ut,GarciaBellido:2008ab}
\begin{equation}\label{eq:Quadpart}
U(\chi)=\frac12\omega^2\chi^2\,, \hspace{20mm} \omega^2 =\frac{ \lambda_{s} M_{\rm P}^2}{3\xi_{s}^2}\,,
\end{equation}
at $M_{\rm P}/\xi_{s}\ll \chi \lesssim \sqrt{3/2}M_\text{P}$, and the usual quartic potential 
\begin{equation}\label{eq:Quartpart}
 V(s)=\frac{\lambda_{s}}{4}s^4\,,
\end{equation}
at $\chi\ll M_P/\xi_s$.
If the non-minimal coupling $\xi_{s}$ is large, the oscillations take place first in the quadratic part and eventually enter the quartic regime, where the Weyl transformation~\eqref{Omega} equals unity. From there on, the dynamical degree of freedom coincides with the $s$ field, and the potential is indeed given by Eq.~\eqref{eq:Quartpart}. However, if the non-minimal coupling $\xi_s$ is small, 
the transition value $\chi\sim M_{\rm P}/\xi_{s}$ is very close to the end of inflation and the reheating stage is essentially dominated by the quartic regime~\eqref{eq:Quartpart} with  $\chi\simeq s$ from the very end of inflation~\cite{Bezrukov:2008ut,GarciaBellido:2008ab}.
Note, however, that if both $\lambda_{hs}$ and $\lambda_{s}$ are very small, the omitted mass terms in Eq.~\eqref{potential} will eventually become dominant. In particular, the inflaton field can relax into a regime governed by the bare mass term $V(s)=m_{s}^2 s^2/2$ before its eventual decay. In the following, we will study the above possibilities case by case in order to fully account for the dynamics in different parts of the parameter space.

\subsubsection{Decay in the quartic part of the potential}

While the inflaton oscillates in the quartic part of its potential, the decay of the $s$ condensate and the associated energy transfer to the SM sector can, in principle, occur by the production of $s$ and $h$ particles,\footnote{As reheating is a non-perturbative process, the $s$ condensate can also fragment into $h$ particles with $m_{h}>m_{s}/2$. For an extended discussion on the topic, see e.g. Refs.~\cite{Kofman:1994rk,Kofman:1997yn,Ichikawa:2008ne,Mukaida:2013xxa,Kainulainen:2016vzv}.} which soon annihilate or decay into other SM particles. Note,  however, that the possibility that the inflaton condensate could dominantly produce stable $s$ particles subsequently reheating the SM sector is ruled out for weakly interacting inflaton-SM sectors since these cannot make the SM sector the dominant energy component prior to big bang nucleosynthesis~\cite{Tenkanen:2016jic}. This imposes an important restriction on the hierarchy of inflaton couplings. In particular, an upper bound on $\lambda_s$  for a given $\lambda_{hs}$ can be inferred from the effective semi-perturbative decay rates of the inflaton condensate~\cite{Ichikawa:2008ne,Nurmi:2015ema,Kainulainen:2016vzv},
\begin{equation}
\label{decayrates}
\Gamma^{(4)}_{s_0\to ss} = 0.023\lambda_{s}^{\frac32}s_0 , \hspace{20mm}
\Gamma^{(4)}_{s_0\to hh} = 0.002\lambda_{hs}^2\lambda_{s}^{-\frac12}s_0 ,
\end{equation}
where the superscripts indicate that the inflaton oscillates in the quartic part of the potential and $s_0(t)=s_{\rm end}\sqrt{t_{\rm end}/t}$ is the time-dependent oscillation amplitude of the inflaton condensate
$s(t)=s_0(t){\rm cn}\left(0.85\sqrt{\lambda_{s}}s_0(t)(t-t_{\rm end}),1/\sqrt{2} \right)\,,$ 
with $s_{\rm end}$ the field value at the end of inflation and cn the elliptic cosine function~\cite{Kofman:1997yn}. For details, see Appendix \ref{decayrates_appendix}.

In the following, we require that the inflaton decay into two $s$ particles produces at most a fraction $C$ of the observed DM abundance $\Omega_{\rm DM}^{\rm obs}$,
\begin{equation}\label{eq:Orest}
 \Omega_{s}\leq C\,\Omega_{\rm DM}^{\,\rm obs}\,,
  \end{equation}
which gives the bound (cf.~Appendix~\ref{appendix} for details)  
  \be
\label{hierarchy}
\lambda_{s} \leq  6\times 10^{-7} \lambda_{hs}^{\frac87}\left(\frac{{\rm GeV}}{m_{s}}\right)^{4/7}\left(\frac{C}{0.1}\right)^{4/7} .
\ee
The somewhat unusual exponents stem from the fractional powers in Eq.~\eqref{decayrates} and the $\sqrt{\lambda_{s}}$ dependence of the time at which the $s$ particles become non-relativistic (i.e. cold DM).  Note that Eq.~\eqref{hierarchy} is more stringent than the one originally found in Ref.~\cite{Tenkanen:2016twd}, as there it was just required that the inflaton decays dominantly into the SM particles, instead of imposing the more restricting condition~\eqref{eq:Orest}.

Since in the present scenario the inflaton decay into SM Higgs particles completely dominates over other processes, one can estimate the reheating temperature by assuming an instantaneous decay at the time at which
\begin{equation}
\Gamma^{(4)}_{s_0\to hh}=H\simeq \frac{\sqrt{V(s_0)}}{3M_\text{P}}\,,\hspace{10mm} \textrm{with} \hspace{10mm} V(s_0)\simeq 
\frac{\lambda_s}{4} s_0^4
\equiv \frac{\pi^2 g_*}{30}  \left(T^{(4)}_{\rm RH}\right)^4\,.
\end{equation}
This corresponds to a reheating temperature~\cite{Tenkanen:2016twd}
\be
\label{RHtemp2}
\frac{T^{(4)}_{\rm RH}}{{\rm GeV}} \simeq 5\times 10^{15}\lambda_{hs}^2\lambda_{s}^{-\frac34} \,.
\ee
Here the superscript $(4)$ indicates again that reheating occurs while the inflaton oscillates in quartic potential and we assumed the number of relativistic degrees of freedom at that epoch to coincide with the SM one, i.e. $g_*=106.75$. 

Several consistency  constraints can be imposed on the reheating temperature~\eqref{RHtemp2}. In particular, in order to ensure that the SM sector achieves thermal equilibrium before the time of freeze-in, we must have 
\be
\label{RHtemplimit}
T^{(4)}_{\rm RH} > 
\begin{cases}	 
m_{s} , \quad m_{s}>\frac12 m_{h}, \\
m_{h} , \quad m_{s}\leq \frac12 m_{h} ,
\end{cases}
\ee
as will become evident in Section~\ref{DMproduction}. Another important constraint comes from imposing that the inflaton potential is not dominated by the bare mass term $\propto m_s^2 s^2$ at the time of reheating, namely 
\begin{equation}
\frac{\lambda_{s}}{4}s_0^4\left(T^{(4)}_{\rm RH}\right)>\frac{m_{s}^2}{2}s_0^2\left(T^{(4)}_{\rm RH}\right)\,.
\end{equation} 
Using $s_0(T^{(4)}_{\rm RH})\simeq 0.007\lambda_{hs}^2/\lambda_{s}M_{\rm P}$~\cite{Tenkanen:2016twd} we obtain
\be
\label{RHtemplimit2}
T^{(4)}_{\rm RH}>\frac12\, \lambda_{s}^{-\frac14}\,m_{s}\,.
\ee
Note that since we are dealing with an essentially quartic scenario in  which the Universe is effectively radiation-dominated from the very end of inflation, the number of inflationary $e$-folds between horizon exit and the end of inflation is independent 
of the reheating temperature~\eqref{RHtemp2}, namely 
\be
\label{Nquartic}
N^{(4)}_* = \ln\left[\left(\frac{\rho_{\rm RH}}{\rho_{\rm end}}\right)^{\frac14}\left(\frac{g_0}{g_*} \right)^{\frac13}\frac{T_0}{T^{(4)}_{\rm RH}}\frac{H_*}{ k_*} \right]\simeq 55 + \frac14\ln\left(\frac{r}{10^{-3}}\right)\,,
\ee
with $\rho_{\rm RH}=\pi^2/30g_*\left(T^{(4)}_{\rm RH}\right)^4$, $\rho_{\rm end}\simeq 3H_*^2M_{\rm P}^2$, $g_0 = 2+21/11\simeq 3.909$ accounting for difference among the neutrino temperature and the present photon temperature $T_0=2.725$ K. For the Hubble parameter at the Planck pivot scale $k_*=0.05\, {\rm Mpc}^{-1}$, we use $H_*=7.84\times 10^{13}\sqrt{r/0.1}$ GeV with the tensor-to-scalar ratio $r=r(\xi_{s}, N_*)$ given by Eq.~\eqref{nsr2app}.  Eq.~\eqref{Nquartic} admits an exact inversion
\be
\label{efolds2}
N^{(4)}_* \simeq  \frac12 {\cal W}\left[\sqrt{\frac{1+6\xi_{s}}{\xi_{s}}} e^{115}\right],
\ee
with ${\cal W}$ the 0-branch of the Lambert function.

\subsubsection{Decay in the quadratic part of the potential}

Let us discuss now the case in which the inflaton couples so weakly to the SM sector that it reaches the quadratic part of its potential  $V(s_0)=m_{s}^2s_0^2/2$ before reheating, i.e. where Eq.~\eqref{RHtemplimit2} is not satisfied. In that case, the  \textit{non-perturbative} inflaton decay rate into Higgs bosons is given by~\cite{Ichikawa:2008ne,Nurmi:2015ema,Kainulainen:2016vzv} 
\begin{equation}
	\Gamma^{(2)}_{s_0\to hh} = \frac{\lambda_{hs}^2s_0^2}{64\pi m_{s}}\sqrt{1-\left(\frac{m_{h}}{m_{s}}\right)^{2}} ,
\end{equation}
with  $m_{s}> m_{h}$.\footnote{We stress that we are considering \textit{particle production in a time-dependent background} which amounts to kinematical condition different from the standard $1\to 2$ particle decay in vacuum (cf. Appendix \ref{decayrates_appendix}). The result means that if Eq.~\eqref{RHtemplimit2} is not satisfied {\it and} $m_{s}\leq m_{h}$, reheating does not occur, light elements do not form, and the scenario is ruled out. Therefore, in this subsection we require the singlet scalar to be more massive than the Higgs (for details, see Refs.~\cite{Ichikawa:2008ne,Nurmi:2015ema,Kainulainen:2016vzv}).} 
Assuming again instantaneous reheating at the time at which  
\begin{equation}
\Gamma^{(2)}_{s_0\to hh}=H= \frac{\sqrt{V(s_0)}}{3M_\text{P}}  \hspace{10mm} \textrm{where now} \hspace{10mm}V(s_0)=\frac 12 m_{s}^2s_0^2\,, 
\end{equation}
we obtain
\be
\label{TRHquadr}
\frac{T^{(2)}_{\rm RH}}{\rm GeV} \simeq 2\times 10^{-9}\lambda_{hs}^{-1}\left(\frac{m_{s}}{\rm GeV}\right)^{\frac32} ,
\ee
where the superscript $(2)$ indicates that reheating occurs while the inflaton oscillates in quadratic low energy potential $V(s_0)=m_{s}^2s_0^2/2$. Requiring again the reheating temperature to be higher than the freeze-in scale, $T^{(2)}_{\rm RH}>m_{s}$, we obtain a bound
\be\label{eq:msGeV}
\frac{m_{s}}{\rm GeV} \geq 2\times 10^{-5}\left(\frac{\lambda_{hs}}{10^{-11}}\right)^2 .
\ee
As in this scenario the bare mass $m_{s}$ is necessarily bigger than $m_{h}$, the condition~\eqref{eq:msGeV} is always satisfied for a Higgs portal coupling $\lambda_{hs}\sim\mathcal{O}\left(10^{-11}\right)$, giving the correct abundance through freeze-in, as we will show in Section~\ref{DMproduction}. We can thus conclude that the model predicts a reheating temperature which, as long as the aforementioned hierarchy requirements are satisfied,  is always above the big bang nucleosynthesis temperature $T_{\rm BBN}\sim 1$~MeV, regardless of the form of the inflaton potential at the time of its decay.

Even though the inflaton condensate cannot decay into $s$ particles while oscillating in the quadratic part of its potential~\cite{Ichikawa:2008ne,Nurmi:2015ema,Kainulainen:2016vzv}, we have to require that the $s$ particles produced in the quartic part do not contribute significantly to the observed DM abundance. In a similar fashion as above, we obtain a bound on the inflaton self-coupling (for details on the derivation, see again Appendix~\ref{appendix}),
\be
\label{hierarchy2}
\lambda_{s} < 2\times 10^{-12}\left(\frac{m_{s}}{\rm GeV}\right)^{\frac23} \left(\frac{C}{0.1}\right)^{\frac23},
\ee
which results from requiring that also in this case the $s$ particles produced during reheating constitute less than a fraction $C$ of the observed DM abundance.

Contrary to the quartic scenario discussed in the previous section, the number of $e$-folds depends now explicitly on the reheating temperature~\eqref{TRHquadr}, namely
\bea
N^{(2)}_* &=& \ln\left[\left(\frac{\rho_{\rm RD}}{\rho_{\rm end}}\right)^{\frac14}\left(\frac{\rho_{\rm RH}}{\rho_{\rm RD}}\right)^{\frac13}\left(\frac{g_0}{g_*} \right)^{\frac13}\frac{T_0}{T^{(2)}_{\rm RH}} \frac{H_* }{ k_*} \right] \\ \nonumber
&\simeq& 55 + \frac{1}{12}\ln\lambda_{s} + \frac13\ln\left(\frac{m_{s}}{T_{\rm RH}}\right) + \frac14\ln\left(\frac{r}{10^{-3}}\right) ,
\eea
where $\rho_{\rm RD}=m_{s}^4/\lambda_{s}$ is the inflaton energy density at the time of the transition from the quartic to the quadratic potential and the other quantities coincide with those given below Eq.~\eqref{Nquartic}. By using again Eqs.~\eqref{nsr2app}, \eqref{xicondition1} and~\eqref{TRHquadr}, we obtain in this case
\be
\label{efolds_quadr}
N_*^{(2)} = \frac{2}{3}{\cal W}\left[\left(\frac{1+6\xi_{s}}{\sqrt{\xi_{s}}}\lambda_{hs}\sqrt{\frac{\rm GeV}{m_{s}}}\right)^{\frac12}e^{94} \right] .
\ee

\subsection{The Palatini case}

So far, the reheating stage  after Palatini inflation has been only studied in Ref.~\cite{Fu:2017iqg}, where the authors studied the self-resonant production of inflaton excitations in a simple $\lambda \phi^4$ theory. However, because in our case the self-coupling $\lambda_{s}$ must be much smaller than the Higgs portal coupling $\lambda_{hs}$ to successfully reheat the SM sector, the following analysis constitutes the first study of the reheating dynamics in Palatini inflation accounting for other production channels beyond the inflaton self-fragmentation.

\begin{figure}[t]
\begin{center} 
\includegraphics[height=0.4\textwidth]{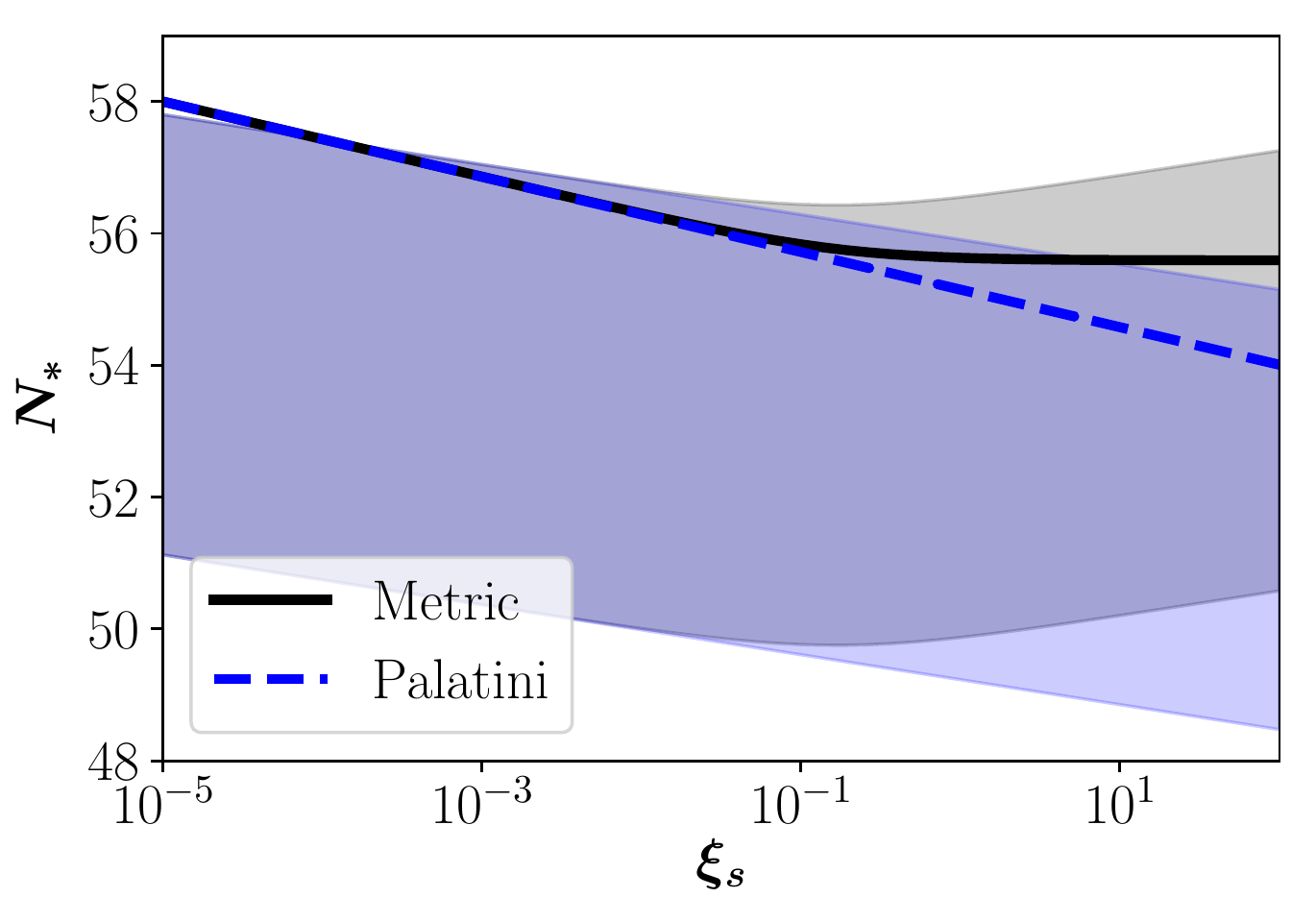} 
\caption{The dependence of the number of $e$-folds $N_*$ on the non-minimal coupling to gravity $\xi_{s}$. The black and blue curves, as well as the grey and blue bands, show the results in metric and Palatini theories, respectively. For details, see the main text.}
\label{fig:RH}
\end{center}
\end{figure}

In the Palatini theory, the inflaton potential during reheating is always quartic to a good accuracy, in contrast with the metric case where it can be quadratic for large $\xi_{s}$ values. The quartic part is indeed reached in less than one $e$-fold after inflation for all the non-minimal couplings of interest, as can be easily verified by numerically solving the equation of motion for $\chi$ after inflation. This allows us to use the decay rates in Eq.~\eqref{decayrates} to find that the upper bound on $\lambda_{s}$ and the reheating temperature, as well as the limits on the latter, do not change as compared to the metric case. 
However,  we can identify small differences in the parametric dependence of the number of $e$-folds $N_*$, both for the case in which reheating occurs in the quartic potential,
\be
\label{efolds3}
N^{(4)}_* = \frac12 {\cal W}\left[\frac{e^{115}}{\sqrt{\xi_{s}}}\right]\,,
\ee
 and for the case in  which the field reaches the low energy quadratic potential $\propto m_s^2s^2$ before decaying,
\be
\label{efolds4}
N_*^{(2)} = \frac{2}{3}{\cal W}\left[\left(\frac{\lambda_{hs}}{\sqrt{\xi_{s}}}\sqrt{\frac{\rm GeV}{m_{s}}}\right)^{\frac12}e^{94} \right] \,.
\ee
Fig.~\ref{fig:RH} presents the dependence of the number of $e$-folds $N_*$ on the non-minimal coupling to gravity $\xi_{s}$.
The black and blue colors correspond respectively to the metric and Palatini cases.
The lines refer to cases where inflation happens in the quartic potential, whereas the bands correspond to the quadratic potential scenarios. The thickness of the bands comes from the allowed values of $m_{s}$ and $\lambda_{hs}$, assuming all DM is produced by the freeze-in mechanism, for a DM mass between 1~keV and 1~PeV (cf. Fig.~\ref{freezeinRH} and Section~\ref{DMfreezein}).

\section{Dark matter production}
\label{DMproduction}

As discussed in Section~\ref{model}, we will consider four distinct scenarios in which the DM component is non-thermally produced. We will start by studying in Section~\ref{DMfreezein} scenarios where reheating produces only a negligible amount of $s$ particles and all the DM is produced by the freeze-in mechanism, and concentrate on the opposite case in Section~\ref{DMreheating}. In Section~\ref{astrophysics}, we will discuss some astrophysical constraints on the model. 
In all cases, we will neglect potential threshold corrections and assume the low energy values of $\lambda_s$ and $\lambda_{hs}$ to coincide with those at the end of the reheating stage. The last assumption is justified by the small renormalization group running expected from the tiny value of these parameters (see below) together with the very restrictive interaction of the hidden sector with the SM particles. All other SM parameters (Higgs self-coupling, $SU(2)$ gauge couplings etc.) do not enter in our tree-level estimates and could, at most, affect the thermalization of the SM plasma after decay of the inflaton condensate.

\subsection{Dark matter from freeze-in}
\label{DMfreezein}

We first assume that DM particles $s$ are produced in the early Universe through 2-to-2 scatterings of SM particles and Higgs decays. If the Higgs portal coupling takes a very small value, $\lambda_{hs}\lesssim 10^{-6}$, the DM sector does not enter into thermal equilibrium with the visible SM sector~\cite{Petraki:2007gq,Enqvist:2014zqa,Kahlhoefer:2018xxo}. In that case, the observed DM relic abundance can be produced by the freeze-in mechanism~\cite{McDonald:2001vt,Hall:2009bx}.
The evolution of the DM number density $n_{s}$ is given by the Boltzmann equation
\be\label{BEFI}
\begin{aligned}
\frac{\td n_{s}}{\td t}+3\,H\,n_{s} 
=& \sum_x \langle \sigma_{x\bar{x}\to ss} v\rangle\, (n_x^{\rm eq})^2 + {\cal C}\, m_{h}\, \Gamma_{h\to ss} \int \frac{\td^3 p_{h}}{(2\pi)^3\,E_{h}} f_{h}^{\rm eq} \,,
\end{aligned}
\ee
where the sum runs over all SM particles, $\langle \sigma_{x\bar{x}\to ss} v\rangle$ corresponds to the thermally averaged annihilation cross-section of the SM species into DM particles, $\Gamma_{h\to ss}$ denotes the Higgs decay width to DM, ${\cal C}\simeq 0.349$~\cite{Frigerio:2011in,Bernal:2018kcw}, and $f_{h}^{\rm eq}$ and $E_{h}$ are the SM Higgs equilibrium distribution function and energy, respectively.
The freeze-in is an infrared process, where essentially all DM is produced at $T\sim m_{h}$ if $m_s\leq m_h/2$, or at $T\sim m_s$ if $m_s> m_h/2$ and not, in the absence of higher-dimensional operators, at the highest temperatures of the Universe (see for instance Ref.~\cite{Hall:2009bx}). Note that in Eq.~\eqref{BEFI}, we are not taking into account possible number-changing DM self-interactions, such as $4s\to 2s$ annihilations. These processes could, in principle, lead to the thermalization of the $s$ particles with themselves even if they do not enter in equilibrium with the SM particles. This can result in a change in the DM abundance even after the initial DM yield has ended~\cite{Carlson:1992fn,Bernal:2015ova,Bernal:2015xba,Heikinheimo:2016yds,Bernal:2017mqb,Heikinheimo:2017ofk,Heeba:2018wtf,Herms:2018ajr,Bernal:2018ins,Bernal:2018hjm}. However, since reheating in our scenario  requires $\lambda_{s} \ll 1$, such processes do not play an important role in the present context and can be therefore safely neglected.

\begin{figure}[t]
\begin{center} 
\includegraphics[height=0.4\textwidth]{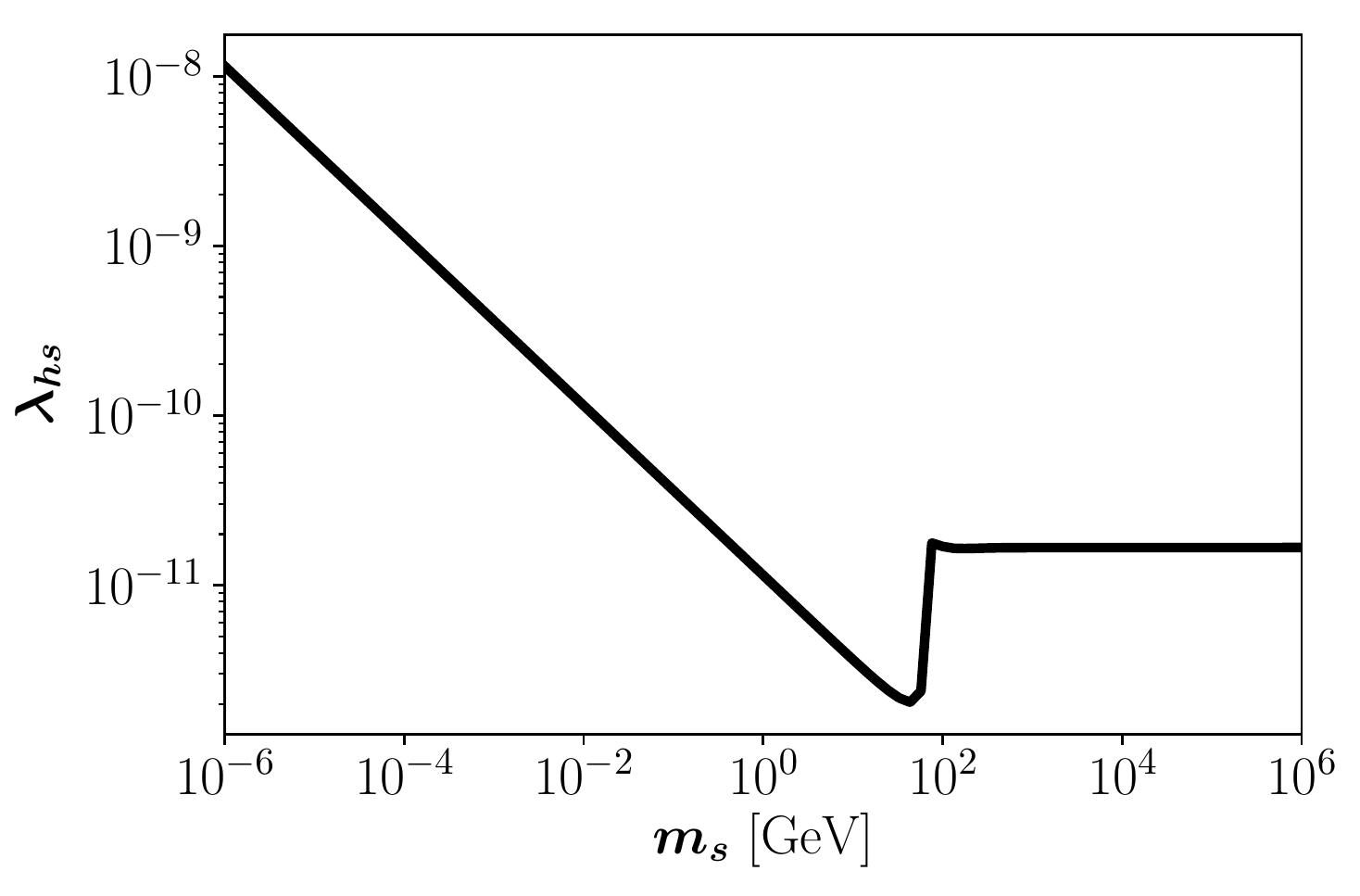} 
\caption{Portal coupling $\lambda_{hs}$ required to reproduce the observed DM relic abundance as a function of the DM mass $m_{s}$, in the freeze-in scenario.}
\label{freezein}
\end{center}
\end{figure}

The DM genesis via the freeze-in mechanism fixes the value of the Higgs portal coupling for each DM mass. The value of the coupling $\lambda_{hs}$ needed to reproduce the observed DM relic abundance for a wide range of DM masses $m_{s}$ varying from the keV to the PeV scale is shown in Fig.~\ref{freezein}.\footnote{To derive this plot we used the {\tt micrOMEGAs5.0} code~\cite{Belanger:2018mqt}.}  Lyman-$\alpha$ forest observations are in tension with a DM mass below a few keV~\cite{Baur:2015jsy,Irsic:2017ixq}, which provides for a natural cut-off for the mass scale.

\begin{figure}[t]
\begin{center} 
\includegraphics[height=0.32\textwidth]{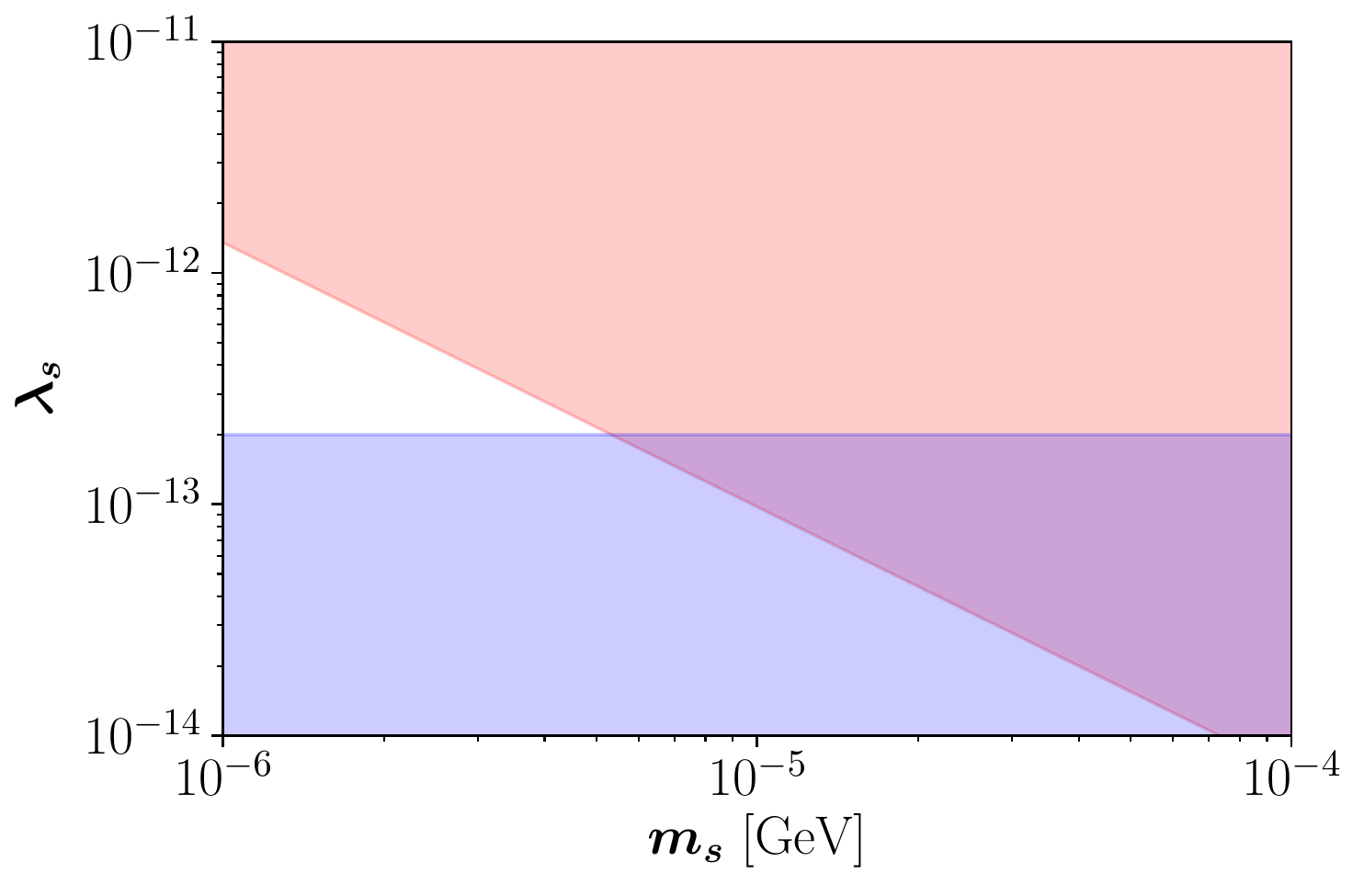} 
\includegraphics[height=0.32\textwidth]{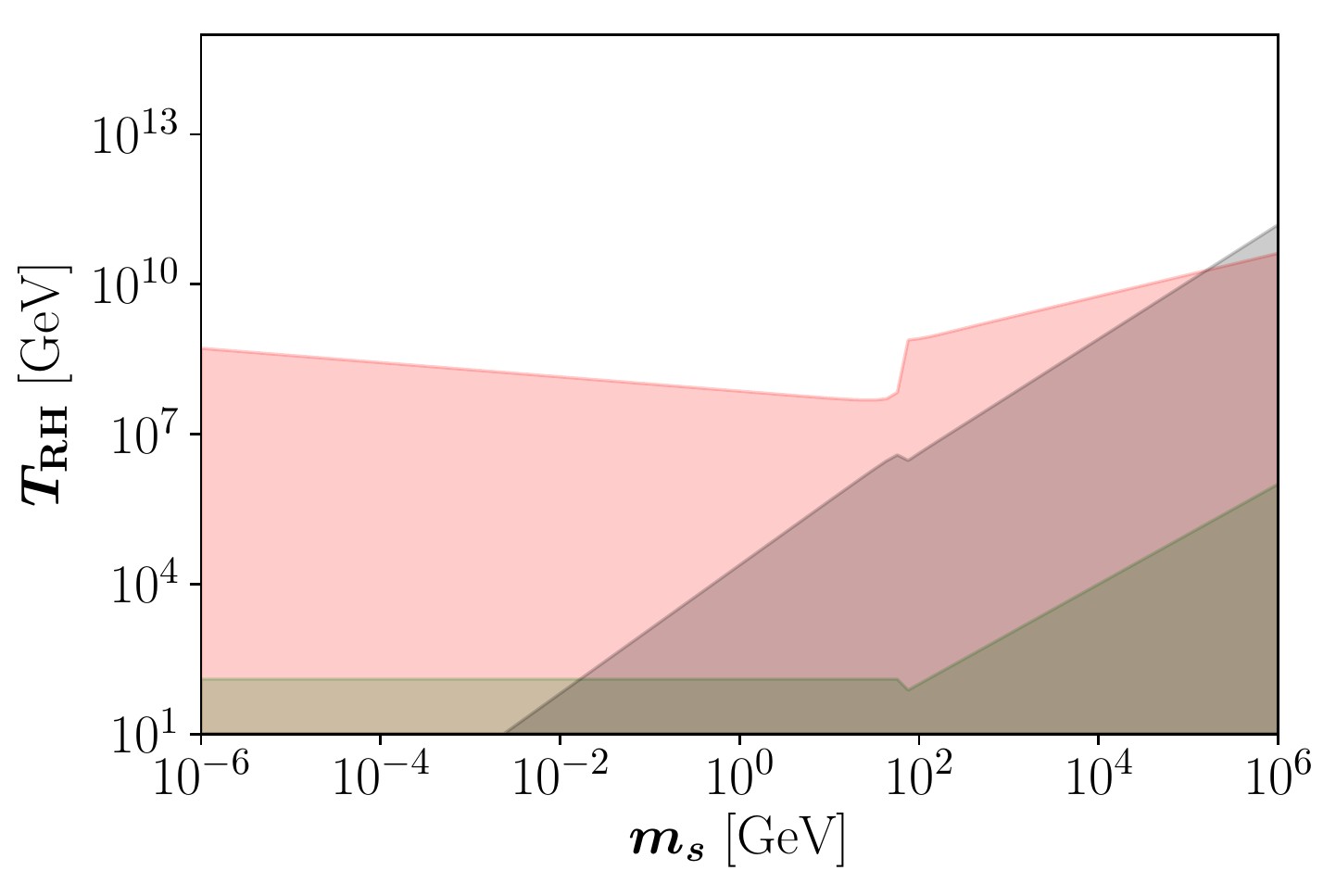} 
\caption{{\bf Scenario 1.} Left panel: Constraint on the inflaton-DM self-interaction $\lambda_{s}$ as a function of the DM mass $m_{s}$, assuming that reheating occurs in \textit{quartic} potential and produces less than $10\%$ of the observed DM abundance.
Right panel: Constraints on the reheating temperature $T_\text{RH}$, assuming all DM was produced by freeze-in. All colored areas are ruled out. The red regions show the bound that the upper bound on $\lambda_{s}$, Eq.~\eqref{hierarchy}, imposes on the reheating temperature~\eqref{RHtemp2}. The green and grey regions are ruled out by Eqs.~\eqref{RHtemplimit} and~\eqref{RHtemplimit2}, respectively. In the blue region $\lambda_s<2\times 10^{-13}$.
}
\label{freezeinRH}
\end{center}
\end{figure}

The left panel of Fig.~\ref{freezeinRH} shows the upper bound on the quartic coupling $\lambda_{s}$ coming from Eq.~\eqref{hierarchy}.
The limit assumes that reheating occurs in a quartic potential and that it produces less than $10\%$ of the observed DM abundance, the rest coming from the freeze-in mechanism. This limit can be translated into a lower bound on the reheating temperature $T_\text{RH}$ (shown in red), coming from Eq.~\eqref{RHtemp2}. Additionally, the green region is also excluded because it leads to a reheating temperature below the freeze-in scale~\eqref{RHtemplimit}, rendering the scenario inconsistent with our assumptions. In the grey area reheating does not happen in a quartic potential, Eq.~\eqref{RHtemplimit2}.
Finally, the blue region, corresponding to $\lambda_s<2\times 10^{-13}$, is incompatible with our inflationary scenario, cf. Fig.~\ref{slow-roll}.

\begin{figure}[t]
\begin{center} 
\includegraphics[height=0.4\textwidth]{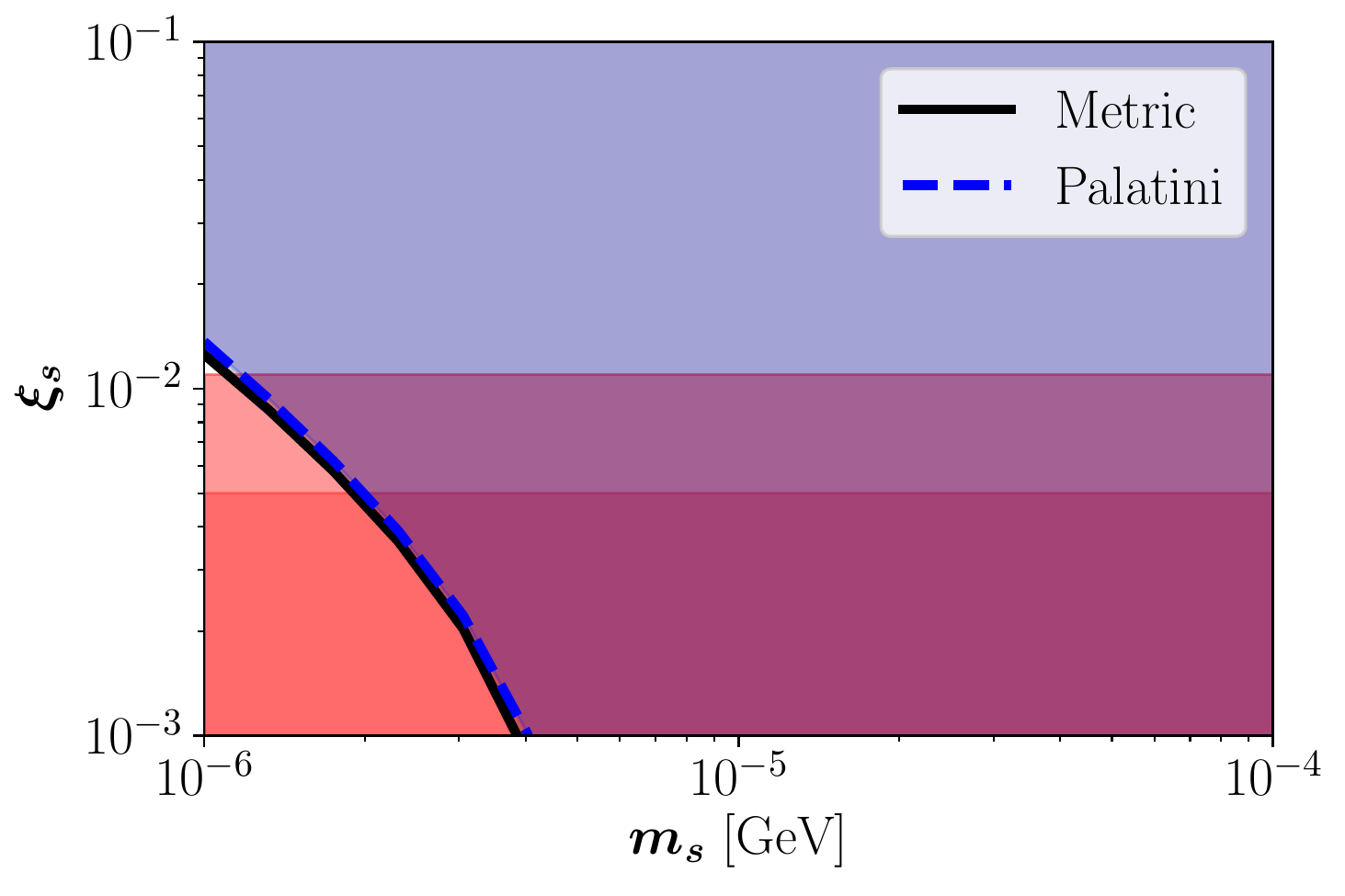} 
\caption{{\bf Scenario 1.} Constraints on the inflaton-DM non-minimal coupling $\xi_{s}$ as a function of DM mass, assuming that reheating occurs in \textit{quartic} potential and produces less than $10\%$ of the observed DM abundance. Here we have used Eq.~\eqref{cobe} to map the constraint on $\lambda_{s}$ to constraint on $\xi_{s}$.
The horizontal red bands correspond to the regions disfavored by Planck at 68$\%$ (light) and 95$\%$~CL (dark).}
\label{fig:TRH}
\end{center}
\end{figure}

Fig.~\ref{fig:TRH} presents the constraints on the inflaton-DM non-minimal coupling $\xi_{s}$, assuming that all the DM component was produced by freeze-in and using Eq.~\eqref{cobe} to map the constraint on $\lambda_{s}$ to a constraint on $\xi_{s}$.
The horizontal red bands correspond to the regions on the $n_S-r$ plane, disfavored by Planck at 68$\%$ (light) and 95$\%$~CL (dark).

\begin{figure}[t]
\begin{center} 
\includegraphics[height=0.32\textwidth]{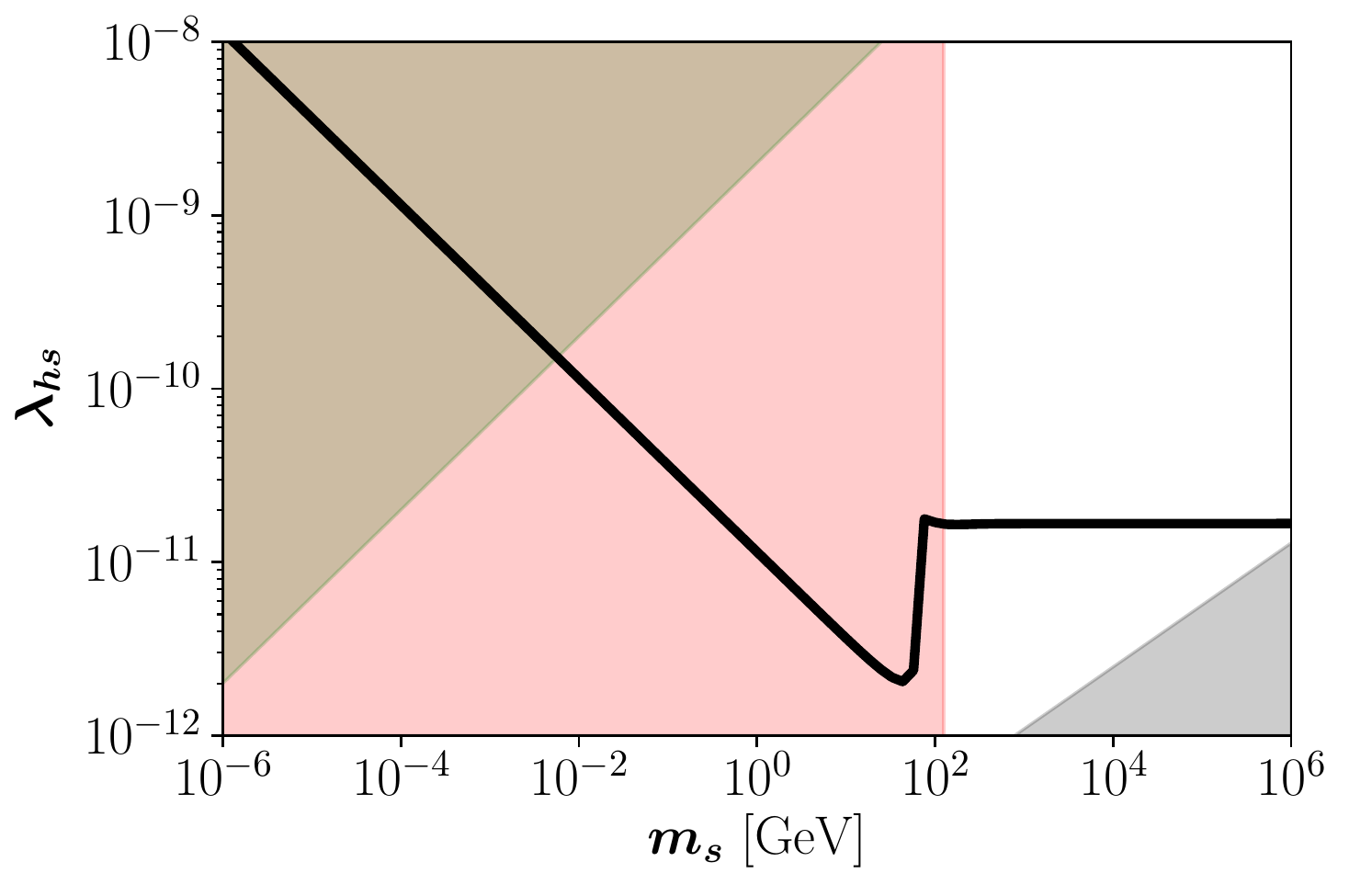} 
\caption{{\bf Scenario 2.} The black thick line corresponds to the values of $\lambda_{hs}$ needed to simultaneously reproduce the observed DM relic abundance via the freeze-in mechanism and have a successful reheating in \textit{quadratic} inflaton potential. All colored areas are ruled out: either because $m_{s}<m_{h}$ (red) or $T_\text{RH}<m_{s}$ (green). In the grey region Eq.~\eqref{RHtemplimit2} is satisfied, leaving the white region available for this scenario. The results show that the scenario is consistent with observations for a broad range of masses, $10^2$~GeV $< m_{s} \lesssim 10^6$~GeV.
}
\label{freezeinRH2a}
\end{center}
\end{figure}

\begin{figure}[t]
\begin{center} 
\includegraphics[height=0.32\textwidth]{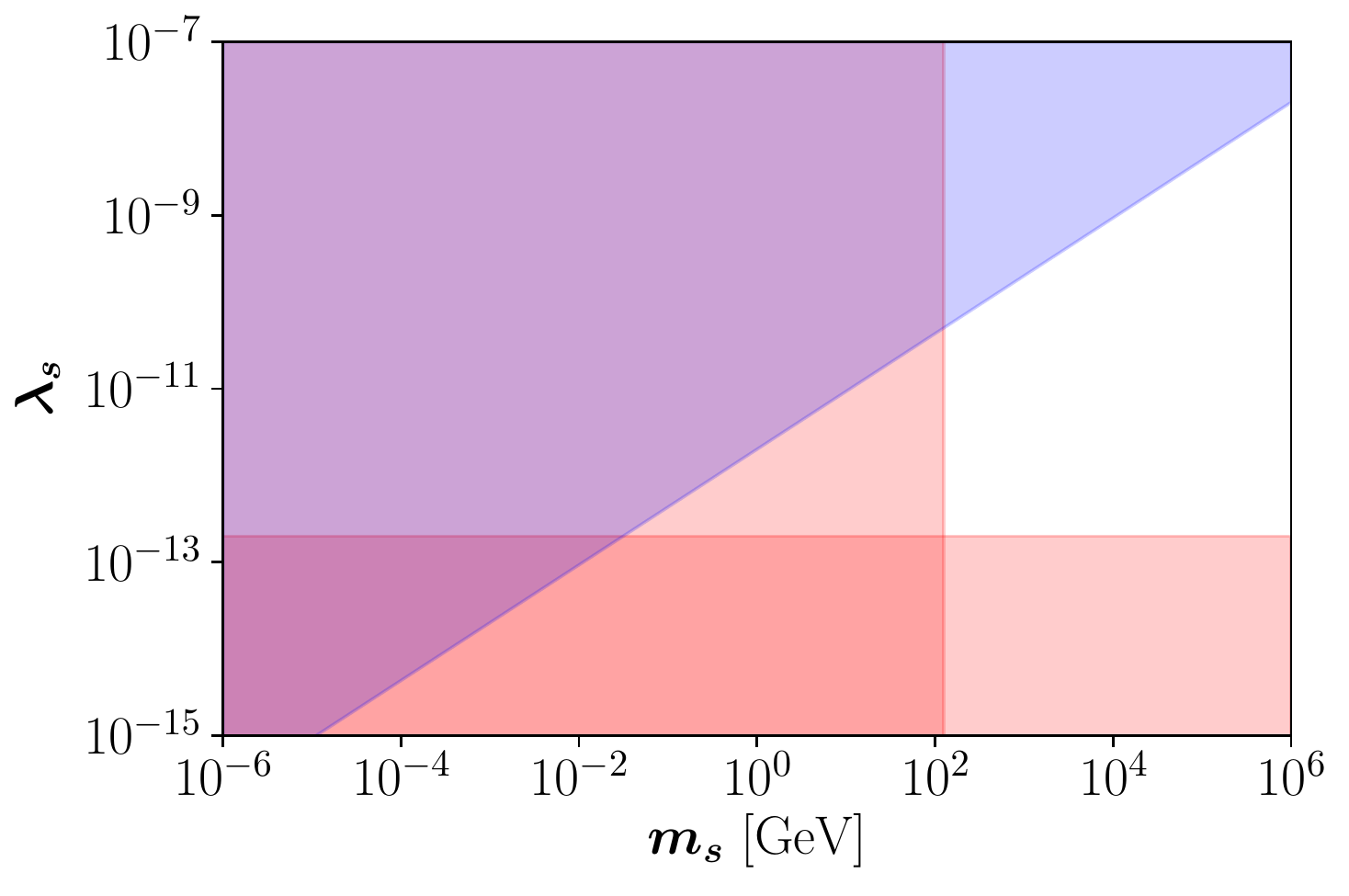} 
\includegraphics[height=0.32\textwidth]{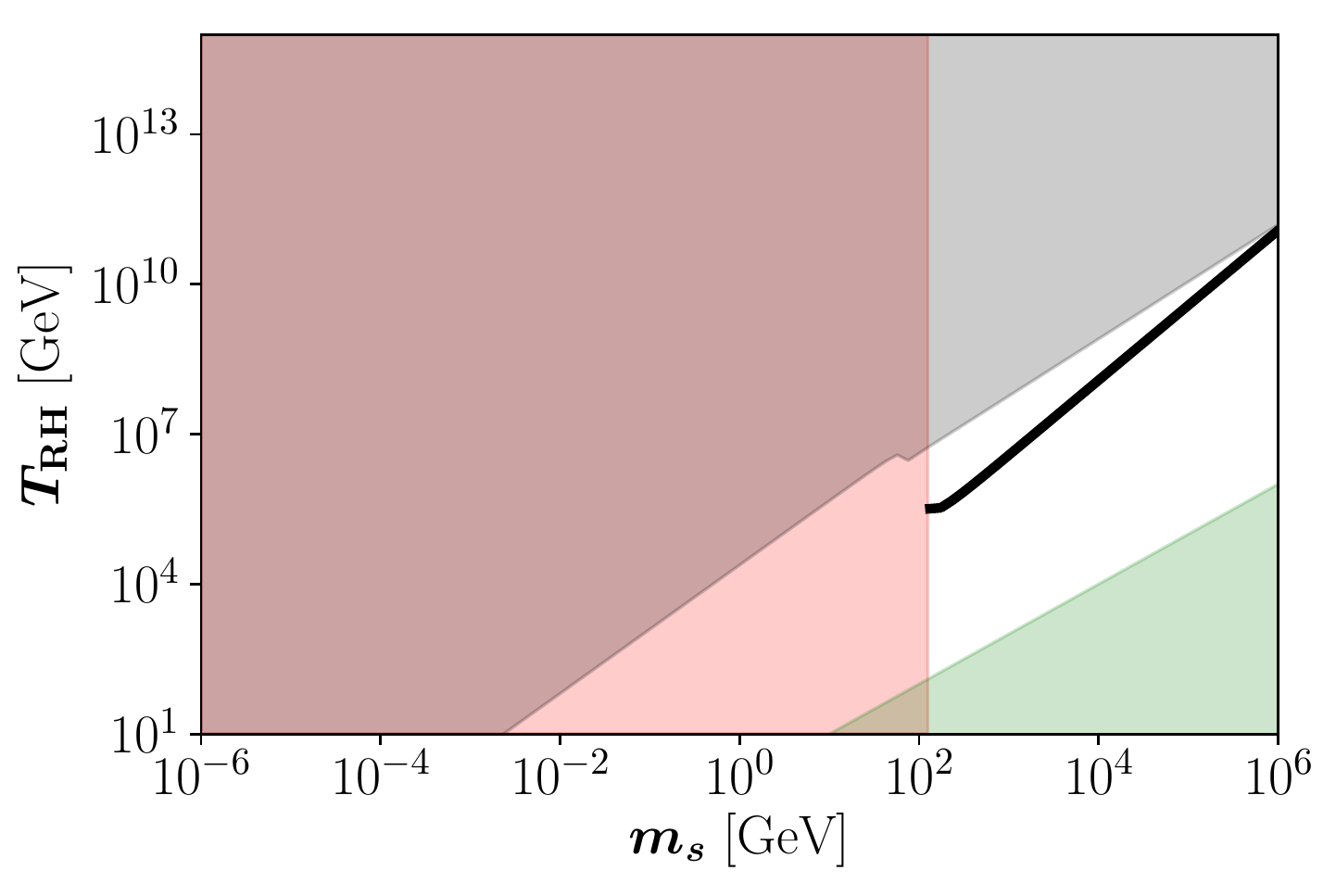} 
\caption{{\bf Scenario 2.} The black thick line corresponds to the values of $T_\text{RH}$ needed to simultaneously reproduce the observed DM relic abundance via the freeze-in mechanism and have a successful reheating in \textit{quadratic} inflaton potential. All colored areas are ruled out: either because $m_{s}<m_{h}$ or $\lambda_s< 2\times 10^{-13}$ (red), or $T_\text{RH}<m_{s}$ (green). In the grey region Eq.~\eqref{RHtemplimit2} is satisfied, whereas in the blue region Eq.~\eqref{hierarchy2} is violated, leaving the white region available for this scenario. The results show that the scenario is consistent with observations for a broad range of masses, $10^2$~GeV$< m_{s} \lesssim 10^6$~GeV.
}
\label{freezeinRH2b}
\end{center}
\end{figure}

\begin{figure}[t]
\begin{center} 
\includegraphics[height=0.4\textwidth]{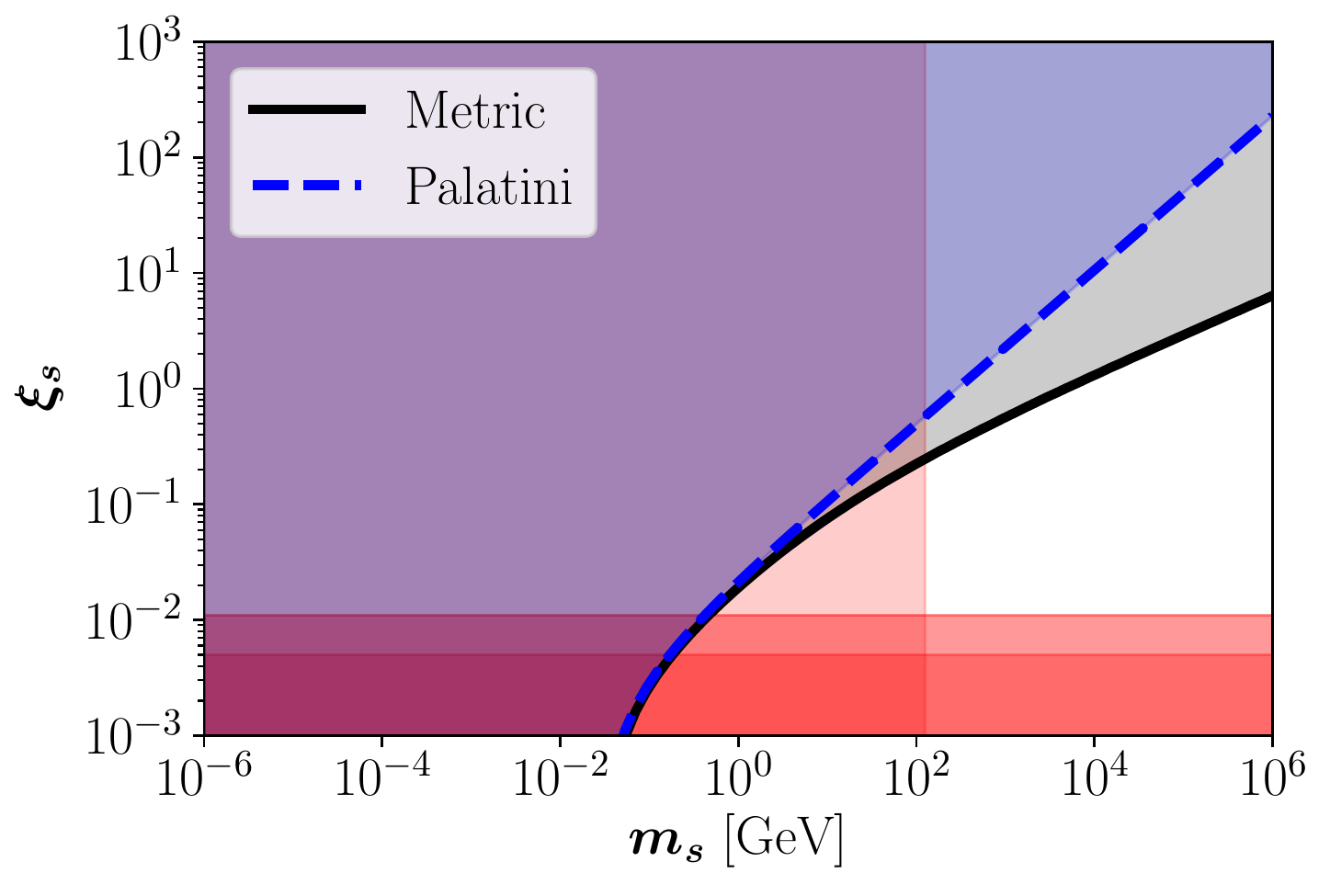} 
\caption{{\bf Scenario 2.} Constraints on the inflaton-DM non-minimal coupling $\xi_{s}$ as a function of DM mass, assuming that the observed DM relic abundance is generated via the freeze-in mechanism and that reheating happens in \textit{quadratic} inflaton potential, for the metric (black) and the Palatini (blue) theories.
In the metric and Palatini theories, non-minimal couplings larger than $\sim 5$ and $\sim 300$ are excluded, respectively.
The white area is allowed in both theories.
Here we have used Eq.~\eqref{cobe} to map the constraint on $\lambda_{s}$ to constraint on $\xi_{s}$.
The horizontal red bands correspond to the regions disfavored by Planck at 68$\%$ (light) and 95$\%$~CL (dark).}
\label{fig:TRH3}
\end{center}
\end{figure}

Figs.~\ref{freezeinRH2a} and~\ref{freezeinRH2b} show the values of $\lambda_{hs}$ and $T_\text{RH}$ needed to simultaneously reproduce the observed DM relic abundance via the freeze-in mechanism and have a successful reheating in a quadratic inflaton potential. All colored areas are ruled out: either because $m_{s}<m_{h}$ or $\lambda_s<2\times 10^{-13}$ (red), or $T_\text{RH}<m_{s}$ (green). In the grey region Eq.~\eqref{RHtemplimit2} is satisfied, whereas in the blue region Eq.~\eqref{hierarchy2} is violated, leaving the white region available for this scenario. The results show that the scenario is consistent with observations for a broad range of masses, $10^2$~GeV $< m_{s} \lesssim 10^6$~GeV. Fig.~\ref{fig:TRH3} shows the constraints on the inflaton-DM non-minimal coupling $\xi_{s}$ as a function of DM mass, for the metric (black) and the Palatini (blue) theories.
Non-minimal couplings larger than $\sim 5$ and $\sim 300$ are excluded in the metric and Palatini cases, respectively.
The white area is allowed in both theories.
In this figure we have used Eq.~\eqref{cobe} to map the constraint on $\lambda_{s}$ to constraint on $\xi_{s}$.
The horizontal red bands correspond to the regions disfavored by Planck at 68$\%$ (light) and 95$\%$~CL (dark).

The above results show that the first scenario in Section~\ref{model}, involving a reheating stage in the quartic part of the potential and a DM  production by freeze-in, is essentially ruled out. On the contrary, the second scenario, involving SM particle production in the low quadratic part of the potential and DM creation by freeze-in after reheating, works well for a broad range of masses $10^2$~GeV$< m_{s} \lesssim 10^6$~GeV, $\lambda_{hs}\sim 10^{-11}$ and $10^5$~GeV$\lesssim T_\text{RH}\lesssim 10^{11}$~GeV. Additionally, the non-minimal coupling is bounded from below $5\times 10^{-3}\lesssim\xi_s$ and has to be smaller than $\sim 3$ or $\sim 200$ in the metric and Palatini theories, respectively.

\subsection{Dark matter from reheating}
\label{DMreheating}

In this scenario, when DM is produced by annihilations of SM particles or by direct decays of the Higgs boson into $s$ particles, the coupling $\lambda_{hs}$ at large masses is always in the $\mathcal{O}(10^{-11})$ ballpark  characteristic for freeze-in scenarios, cf. Fig.~\ref{freezein}. However, even if the Higgs portal coupling  is smaller than the value required for the freeze-in mechanism to  work, the model under consideration could  still produce the observed DM during the reheating stage. In particular, it may happen that the inflaton transfers most of its energy density into the SM sector but simultaneously produces an amount of $s$ particles leading to the observed DM abundance. Quantitatively, this is the case when the branching ratio in Eq.~\eqref{br} is very small and the factor $C$ in Eqs.~\eqref{hierarchy} or~\eqref{hierarchy2} is very close to unity. In the following, we will first analyze the scenario where reheating occurs in the quartic part of the potential and then the case in which it occurs in the low-energy quadratic part. These correspond to the scenarios 3 and 4 discussed in Section \ref{model}.

\subsubsection{The quartic case}

In this case, the correct DM abundance is obtained from Eq.~\eqref{hierarchy} with $C=1$,
\be
\label{lambdaDM}
\lambda_{s} = 2\times 10^{-6}\left(\lambda_{hs}^2\frac{{\rm GeV}}{m_{s}}\right)^{\frac{4}{7}} ,
\ee
provided that 
\be\label{constraint-DM-RH}
\frac{m_{s}}{\rm GeV} \gg 10^{-12}\left(\frac{\lambda_{hs}}{10^{-12}}\right)^{\frac14}.
\ee
This ensures a small branching ratio ${\rm BR}\ll 1$ (see Eq. \eqref{br}) while simultaneously giving the correct DM abundance. Requiring additionally that the reheating temperature is high enough for the inflaton to decay in the quartic potential, we get (cf. Eq.~\eqref{RHtemplimit2})
\be\label{constraint-DM-RH2}
\frac{m_{s}}{\rm GeV} < 2\times 10^{2}\left(\frac{\lambda_{hs}}{10^{-12}}\right)^2 .
\ee
The scenario requires that $\lambda_{hs}$ is small enough as to not contribute to the total DM abundance in significant amounts, as can be inferred from Eq.~\eqref{BEFI}. In the following, we require that freeze-in produces less than $10\%$ of the observed DM abundance.

\begin{figure}[t]
\begin{center} 
\includegraphics[height=0.32\textwidth]{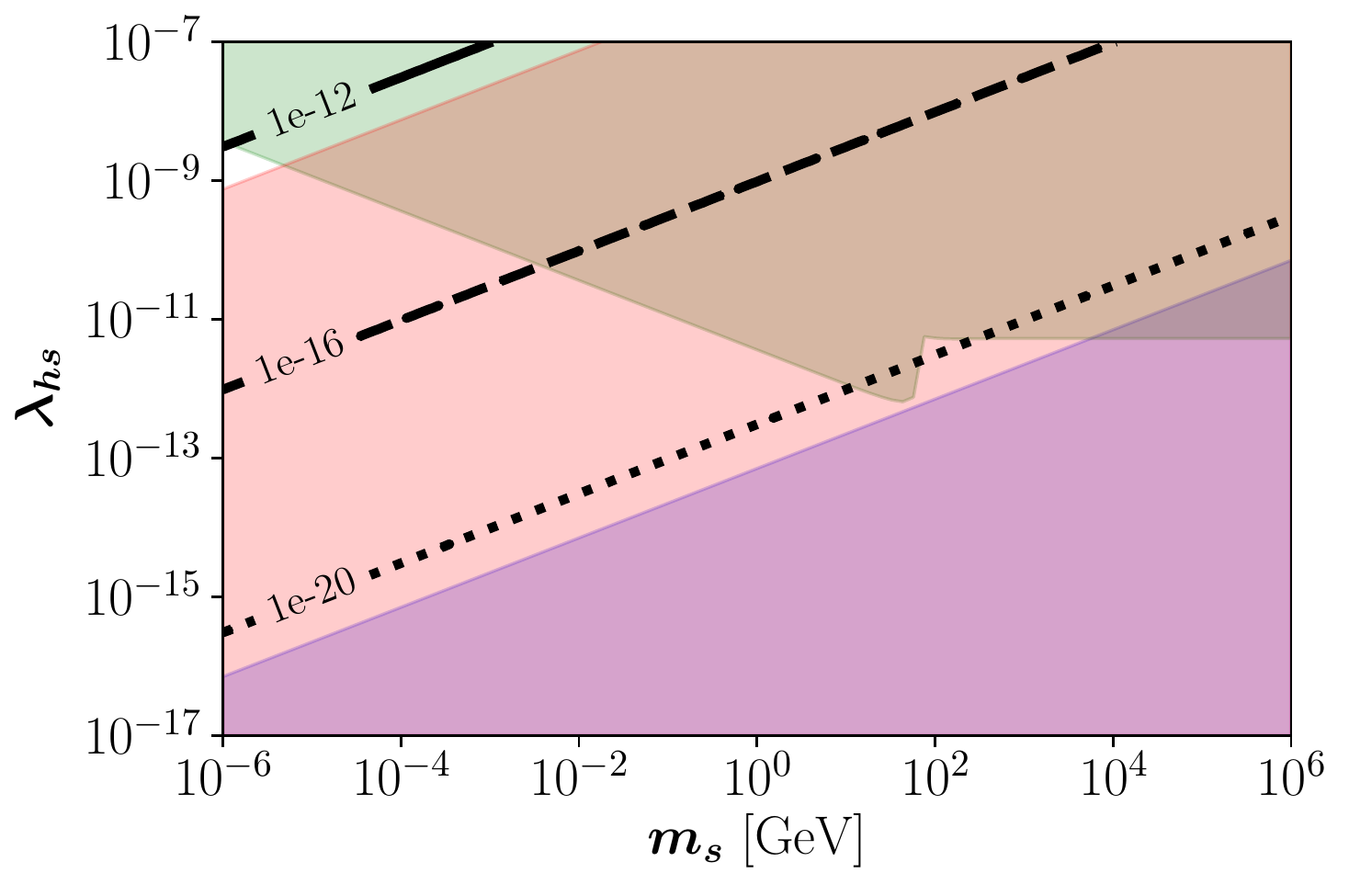} 
	\caption{{\bf Scenario 3.} Contours for the quartic coupling $\lambda_s$ needed to generate the DM observed abundance from reheating in a \textit{quartic} potential, for $\lambda_s=10^{-12}$ (solid line), $10^{-16}$ (dashed line) and $10^{-20}$ (dotted line).
	In the green region the freeze-in mechanism produces more than 10$\%$ of the total DM abundance.
	The blue region violates the constraint from Eq.~\eqref{constraint-DM-RH2}. The red region corresponds to values $\lambda_{s}\lesssim 2\times 10^{-13}$, incompatible with our inflationary scenario.
}
\label{fig:DM-RH}
\end{center}
\end{figure}

Fig.~\ref{fig:DM-RH} shows contours for the quartic coupling $\lambda_s$ required to generate the observed DM abundance from reheating, taking $\lambda_{s}=10^{-12}$ (solid line), $10^{-16}$ (dashed line) and $10^{-20}$ (dotted line).
In the green region a significant DM component is produced by the freeze-in mechanism. The grey region violates the constraint from Eq.~\eqref{constraint-DM-RH}. The red region corresponds to values $\lambda_{s}\lesssim 2\times 10^{-13}$ incompatible with our inflationary scenario (cf. Fig.~\ref{slow-roll}). The results show that this case, corresponding to the third one in Section~\ref{model}, is only marginally allowed for $m_{s}=\mathcal{O}(1)$~keV and $\lambda_{hs}\simeq 10^{-9}$.

\subsubsection{The quadratic case}

If  the inflaton reaches the quadratic part of its potential before decaying into the SM particles, the correct DM abundance, consisting of particles that were produced while the inflaton was still oscillating in quartic potential, is obtained from Eq.~\eqref{hierarchy2} with $C=1$,
\be
\lambda_{s} = 10^{-12}\left(\frac{m_{s}}{\rm GeV}\right)^{\frac{2}{3}} ,
\ee
provided that
\be
\label{constraint-DM-RH3}
\frac{m_{s}}{\rm GeV} > 0.3\left(\frac{\lambda_{hs}}{10^{-11}}\right)^2 ,
\ee\
and
\be
\label{constraint-DM-RH4}
\lambda_{s} < 0.3\,\lambda_{hs}\,.
\ee
This ensures that the inflaton decays dominantly into the SM sector only after its transition to the quadratic part of the potential, while simultaneously producing enough $s$ particles in the quartic region to constitute the observed DM abundance. Since in this case the mass of the $s$ particles must exceed the Higgs mass for successful reheating ($m_{s}>m_{h}$), all values of $\lambda_{hs}$ giving a negligible contribution to the DM abundance via freeze-in allow $m_{s}$ to take values leading to the right DM abundance, as can be seen from Eq.~\eqref{constraint-DM-RH3}.

\begin{figure}[t]
\begin{center} 
\includegraphics[height=0.32\textwidth]{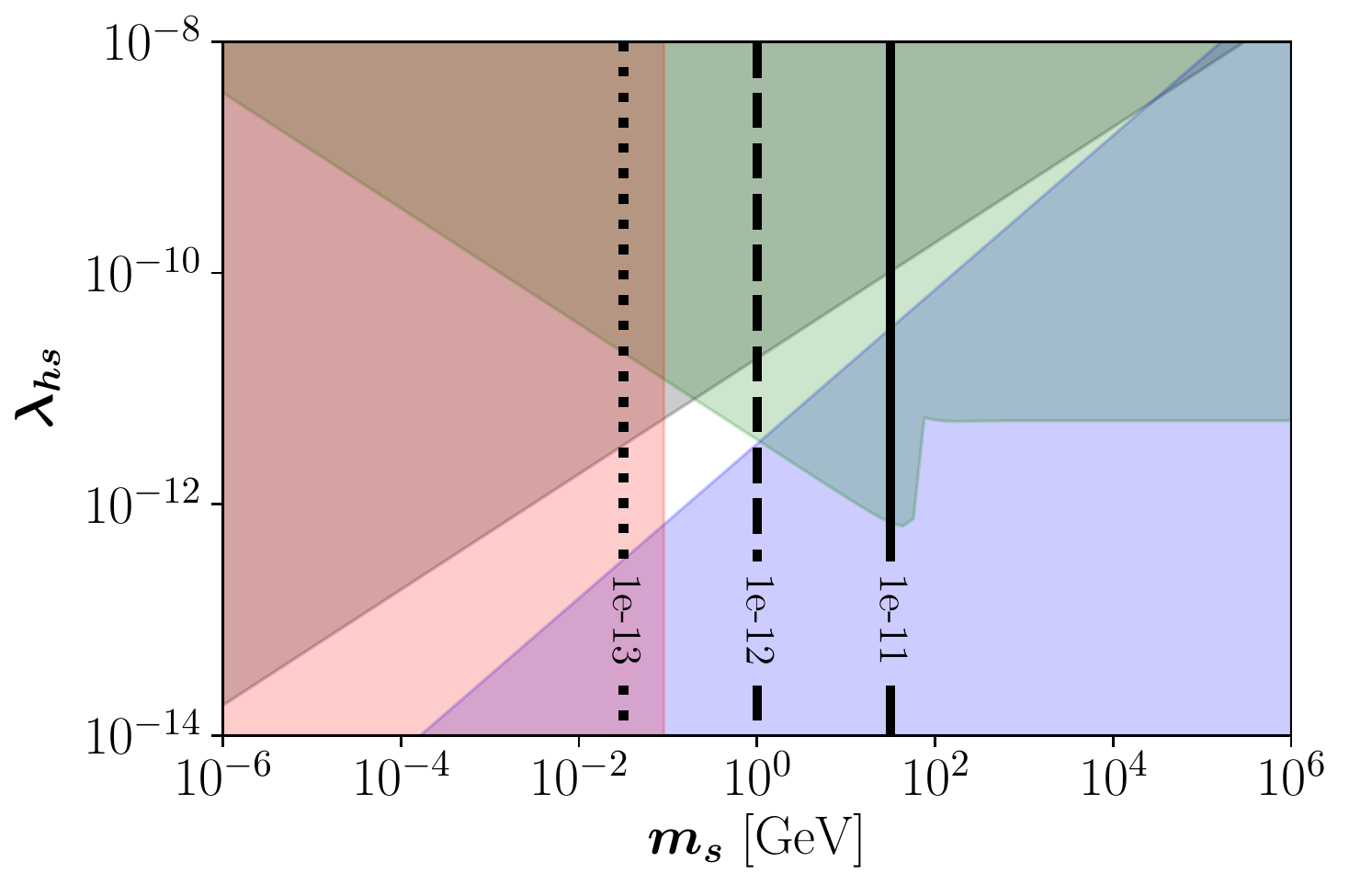} 
\includegraphics[height=0.32\textwidth]{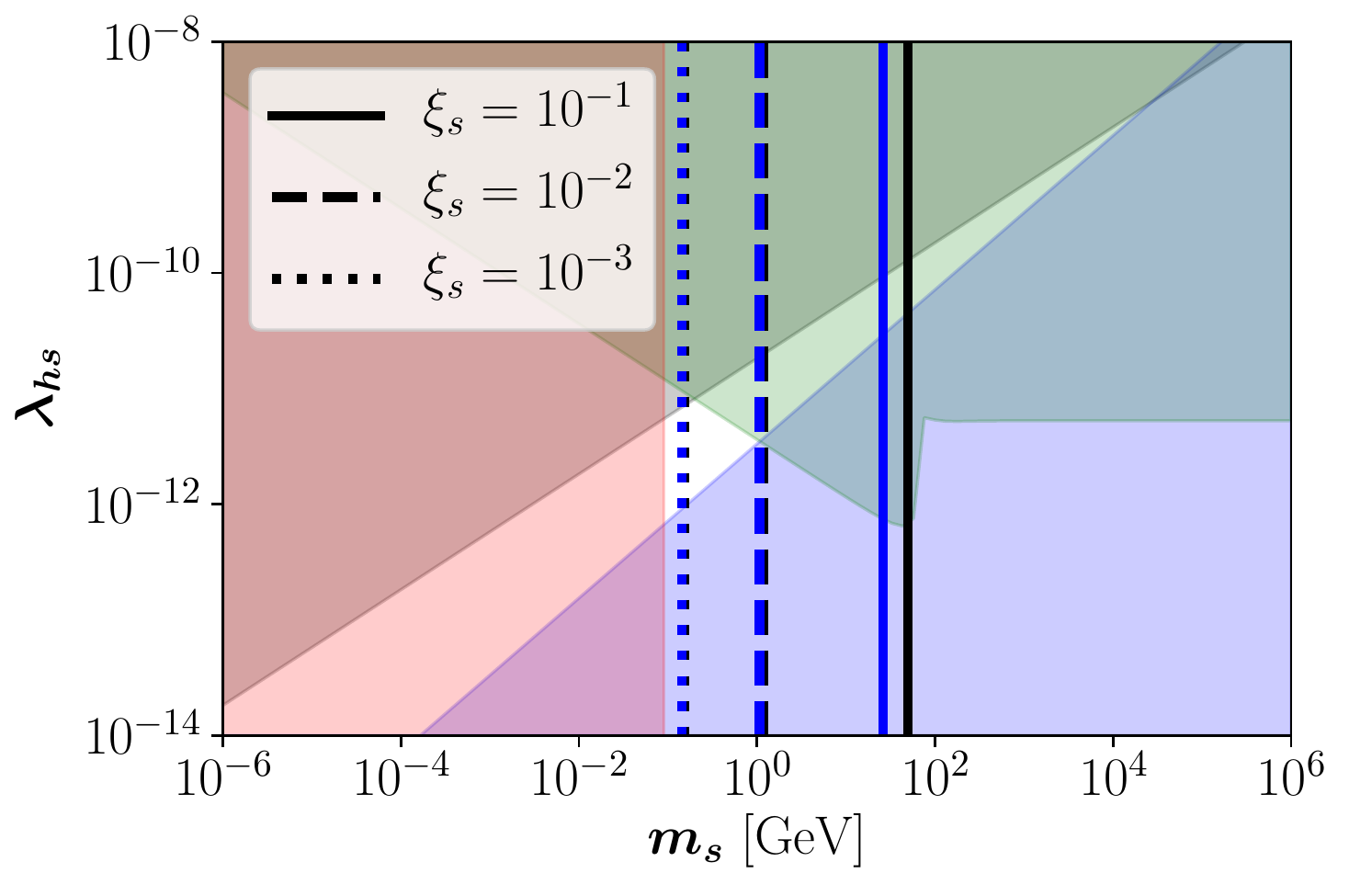} 
	\caption{{\bf Scenario 4.} Contours for the quartic coupling $\lambda_{s}$ (left panel) and non-minimal coupling $\xi_{s}$ (right panel) needed to generate the observed DM abundance from reheating in a \textit{quadratic} potential.
	In the left panel, $\lambda_{s}=10^{-11}$ (solid line), $10^{-12}$ (dashed line), and $10^{-13}$ (dotted line). In the right panel, $\xi_{s}=10^{-1}$ (solid lines), $10^{-2}$ (dashed lines), and $10^{-3}$ (dotted lines), for the metric (black lines) and Palatini (blue lines) theories. 
	In the green region the freeze-in mechanism produces more than 10$\%$ of the total DM abundance.
	The grey region in the upper left corner violates the constraint from Eq.~\eqref{constraint-DM-RH3}, whereas the blue region in the lower right corner violates the constraint from Eq.~\eqref{constraint-DM-RH4}.
	The red region corresponds to values $\lambda_s\lesssim 2\times 10^{-13}$, incompatible with our inflationary scenario. The white region is allowed, showing that the scenario is only viable for $m_s\sim \mathcal{O}(100)$~MeV and $\lambda_{hs}\sim 10^{-12}$.
}
\label{fig:DM-RH3}
\end{center}
\end{figure}

Fig.~\ref{fig:DM-RH3} displays contours for the values of the quartic coupling $\lambda_{s}$ (left panel) and the non-minimal coupling $\xi_{s}$ (right panel) needed to generate the DM observed abundance from reheating in a quadratic potential.
In the left panel, $\lambda_{s}=10^{-11}$ (solid line), $10^{-12}$ (dashed line), and $10^{-13}$ (dotted line). In the right panel, $\xi_{s}=10^{-1}$ (solid lines), $10^{-2}$ (dashed lines) and $10^{-3}$ (dotted lines), for the metric (black lines) and the Palatini (blue lines) scenarios.
In the upper green region the freeze-in mechanism produces more than 10$\%$ of the total DM abundance. The grey region in the upper left corner violates the constraint from Eq.~\eqref{constraint-DM-RH3}, whereas the blue region in the lower right corner violates the constraint from Eq.~\eqref{constraint-DM-RH4}. The red region on the left corresponds to values $\lambda_{s}\lesssim 2\times 10^{-13}$, incompatible with our inflationary scenario (see again Fig.~\ref{slow-roll}).

The above results show that the third scenario in Section~\ref{model} involving a reheating stage in the quartic part of the potential and a DM production during reheating is marginally allowed for DM masses in the keV ballpark and $\lambda_{hs}\sim 10^{-9}$. A similar thing happens in the fourth scenario, involving SM particle production in the low quadratic part of the inflaton potential and DM creation during reheating, which is only viable for $m_s\sim\mathcal{O}(100)$~MeV and $\lambda_{hs}\sim 10^{-12}$.

\subsection{Astrophysical constraints}
\label{astrophysics}

The detection perspectives for the present model at colliders and at indirect or direct detection experiments are challenging, as it is typically the case of FIMP DM scenarios.\footnote{There are, however, FIMP scenarios testable with standard searches, see e.g. Refs.~\cite{Heikinheimo:2018duk,Hambye:2018dpi,Belanger:2018sti}.} However, on top of the cosmological constraints on inflation discussed above, there are some important constraints coming from astrophysics, showing that the present scenario is clearly falsifiable.

For a singlet scalar, the DM-DM elastic scattering strength is given by~\cite{Heikinheimo:2016yds}
\be
\frac{\sigma_{s}}{m_{s}} \simeq \frac{9\lambda_{s}^2}{32\pi m_{s}^3} ,
\ee
which applies when the particles are very light $m_{s}\ll m_{h}$ or very feebly coupled to the Higgs. These self-scatterings due to the quartic coupling are velocity-independent and must be compared with galaxy cluster bounds, like the one following from the bullet cluster observation, namely $\sigma_s/m_s<1.25$~cm$^2$/g at $68\%$ CL~\cite{Markevitch:2003at,Clowe:2003tk,Randall:2007ph}. This bound is naturally satisfied in our framework since the reheating constraints demand a coupling hierarchy $\lambda_{s}<\lambda_{hs}$. 

Two long-standing puzzles of the collisionless cold DM paradigm are the `cusp vs. core'~\cite{Moore:1994yx, Flores:1994gz, Navarro:1996gj, deBlok:2009sp, Oh:2010mc, Walker:2011zu} and the `too-big-to-fail'~\cite{BoylanKolchin:2011de, Garrison-Kimmel:2014vqa} problems. These issues ---collectively referred to as \textit{small scale structure problems} of the $\Lambda$CDM model~\cite{Tulin:2017ara}--- could be alleviated by DM self-interactions if the associated cross section at the scale of dwarf galaxies is in the range $0.1\lesssim\sigma_{s}/m_{s}\lesssim 10$~cm$^2$/g~\cite{Spergel:1999mh, Wandelt:2000ad, Buckley:2009in, Vogelsberger:2012ku, Rocha:2012jg, Peter:2012jh, Zavala:2012us, Vogelsberger:2014pda, Elbert:2014bma, Kaplinghat:2015aga}. Unfortunately, due to the coupling hierarchy demanded by the reheating dynamics, the scalar DM considered in this paper cannot be a solution of the small scale structure problems. This means that in case of positive observation of sizable $\sigma_{s}/m_{s}$, all the (otherwise allowed) scenarios described in this paper would be ruled out.

 Certain subclasses of the present model naturally favor DM in the keV ballpark. We note that this is precisely the mass range where annihilating or decaying DM may provide a viable explanation of the claimed observation of an $X$-ray line at 3.5~keV in various astrophysical systems~\cite{Bulbul:2014sua,Boyarsky:2014jta}; see Refs.~\cite{Merle:2014xpa,Heikinheimo:2016yds,Heeck:2017xbu,Brdar:2017wgy,Kahlhoefer:2018xxo} for different ways of connecting this line to the freeze-in mechanism. In particular, as recently discussed in Ref.~\cite{Kahlhoefer:2018xxo}, if the assumed $\mathbb{Z}_2$ symmetry is broken, already a singlet scalar may yield the observed signal while simultaneously maintaining a long enough lifetime to act as DM. Such scenarios may be testable with future $X$-ray missions. 
\section{Conclusions}
\label{conclusions}

In this paper we have studied a set of minimalistic scenarios in which a real singlet scalar field $s$  drives inflation, reheats the Universe after inflation, and constitutes all the observed DM component. We considered two benchmark models where the DM abundance is generated either by production during reheating or via non-thermal freeze-in, and showed that each of these cases can be further divided into two subclasses. We performed a detailed analysis of inflationary dynamics including the two scalar fields present in the model (the $s$ inflaton-DM field and the SM Higgs), and determined the relevant observables in two different, but well-motivated, theories of gravity (metric and Palatini). We showed that both cases are in very good agreement with observations, provided that the inflaton-DM self-interaction coupling satisfies $\lambda_{s}\gtrsim 2\times 10^{-13}$ and that the $s$-field is mildly coupled to gravity. 

When computing the DM abundance, we took into account all relevant DM production channels in the keV to PeV mass range and found that, for certain initial conditions for inflation and values of the couplings (cf. scenario 2 in Section \ref{model}), the scenario works well for large DM masses, $10^2$~GeV $\lesssim m_{s}\lesssim 10^6$~GeV. We also found that a certain subclass of the model allows for a small mass window at the keV scale (cf. scenario 3 in Section \ref{model}), with potentially interesting observational consequences. A third allowed mass window is at $\mathcal{O}(100)$ MeV (cf. scenario 4 in Section \ref{model}) but in this case testability of the scenario is very limited. We also showed that a potential detection of a sizable DM self-interaction cross-section per DM particle mass could not be explained within the present model and would therefore rule out all the (otherwise allowed) scenarios studied in this paper.

The inflaton-DM model studied in this paper is not only a very economical model for explaining both inflation and DM, but also constitutes an interesting example of a scenario where tiny couplings to the SM sector can be constrained by carefully investigating the inflationary and post-inflationary dynamics. The comparison of the detailed predictions of the model with cosmological and astrophysical observables sheds light on particle physics scenarios which would remain otherwise unconstrained or untestable. \\

\noindent {\bf Note added:} On the day this paper was announced, another paper studying a model  similar to ours but focused on the stability of the electroweak (EW) vacuum appeared on the arXiv repository~\cite{Rusak:2018kel}. This article concluded that in order to ensure the absolute stability a hierarchy $\lambda_{hs}/\xi_s \simeq 10^{-10}$ between the inflaton couplings to the Higgs field and gravity is required. This ratio is in slight tension with the one needed for a feebly-coupled scalar field to drive inflation and later constitute the whole DM component. We note, however, that the stability of the Standard Model vacuum is still an open issue given the experimental and theoretical uncertainties on the relation between the Monte Carlo top quark mass measured at collider experiments and the top Yukawa coupling entering in the SM renormalization group equations (see for instance Ref.~\cite{Bezrukov:2014ina} and references therein). For this reason we have decided not to include the above bound in our analysis.

\acknowledgments
We thank N. Arkani-Hamed, M. Heikinheimo, M. Hertzberg, R. Jinno, M. Postma, S. Rusak, V. Vaskonen, L.-P. Wahlman, and especially T. Takahashi for correspondence and discussions. We also thank an anonymous referee for valuable comments on the manuscript. J.R. and T.T. thank Universidad Antonio Nari\~{n}o for hospitality.  J.R.  was supported by the DFG through the project TRR33 `The Dark Universe' and by the DAAD PPP Kolumbien 2018 project 57394560/ 
COLCIENCIAS - DAAD grant 110278258747. T.T. is supported by the Simons foundation and the U.K. Science and Technology Facilities Council grant ST/J001546/1. N.B. is partially supported by Spanish MINECO under Grant FPA2017-84543-P. This project has received funding from the European Union's Horizon 2020 research and innovation programme under the Marie Sklodowska-Curie grant agreements 674896 and 690575, from COLCIENCIAS grant number 123365843539 RC FP44842-081-2014, and from Universidad Antonio Nari\~no grants 2017239 and 2018204. In addition to the software packages cited above, this research made use of IPython~\cite{Perez:2007emg}, Matplotlib~\cite{Hunter:2007ouj}, and SciPy~\cite{SciPy}.

\appendix


\section{Multifield inflation}
\label{multifield_appendix}

In this appendix we extend the discussion in Sec.~\ref{inflation} to the full two-field case. While the scenario at hand simplifies to a single-field model for $h=0$, the resulting framework in the multifield case emphasizes several features of the model that were not properly identified in previous analytical studies and sheds some light on the results of numerical simulations on multifield inflation~\cite{Carrilho:2018ffi}. In particular, we show that for generic values of the non-minimal couplings, both the $h$ and $s$ fields  participate in inflation and none of them should be understood as a spectator field. Interestingly, in spite of dealing with an \textit{intrinsically multifield inflationary model}, no isocurvature perturbations nor non-Gaussianities are produced during the inflationary stage.
To see this explicitly let us consider the graviscalar part of the DM-driven inflationary action at  field values $h,\,s\gg \mu_s,\,\mu_h$, namely
\be \label{nonminimal_action}
	S_J = \int d^4x \sqrt{-g}\left[\frac{f(s,h)}{2} g^{\mu\nu}R_{\mu\nu}(\Gamma) - \frac{1}{2} g^{\mu\nu}\partial_{\mu}s\partial_{\nu}s- \frac{1}{2} g^{\mu\nu}\partial_{\mu}h\partial_{\nu}h - V(s) \right] \,,
\ee
with 
\be
f(s,h)\equiv M_{\rm P}^2 + \xi_{s} s^2+\xi_{h} h^2\,, \hspace{10mm} V(s,h)=\frac{\lambda_{s}}{4} s^4 +\frac{\lambda_{h}}{4}h^4+\frac{\lambda_{hs}}{4}h^2 s^2 \,.
\ee 
Note that for $ \xi_{s} s^2+\xi_h h^2 \gg M_\text{P}^2$, the above expression develops an internal dilatation symmetry. This means that the Goldstone theorem is applicable and, consequently, it is always possible to identify a dilaton field $\Phi$ displaying only derivative couplings to matter~\cite{GarciaBellido:2011de,Casas:2018fum}. As we will see below, this emergent symmetry has far-reaching consequences.

The cosmological implications of Eq.~\eqref{nonminimal_action} are again more easily understood in the  Einstein frame, in which the non-linearities associated with the non-trivial kinetic mixing among the scalar fields and the metric are transferred to the scalar sector of the theory. This frame is achieved by performing a Weyl transformation $g_{\mu\nu} \to \Omega^2(s,\,h)\,g_{\mu\nu}$ with 
\be \label{Omega1}
 \Omega^2(s,\,h)\equiv 1+\frac{\xi_{s} s^2}{M_{\rm P}^2}+\frac{\xi_{h} h^2}{M_{\rm P}^2} \,.
\ee
Again, while in GR the equation of motion for the connection renders the two formalisms equivalent, this is not true in the presence of non-minimal couplings~\cite{Sotiriou:2008rp}. This fact is indeed reflected in the Einstein-frame action
\be \label{einsteinframe}
	S_E = \int d^4x \sqrt{-g}\left[\frac{M_{\rm P}^2}{2} R - 
\frac{1}{2} g^{\mu\nu} \gamma_{ab} \partial_\mu \varphi^a \partial_\nu \varphi^b - U(\varphi^a)\right] \,,
\ee
where $\varphi^{a}\equiv (\varphi^1,\varphi^2)=(s,h)$,  $U(\varphi^a)\equiv V(s)/ \Omega^{4}(s,h)$ is the Weyl-rescaled potential, $R = g^{\mu\nu}R_{\mu\nu}(\Gamma_E)$ and
\be
  \gamma_{ab} = \frac{1}{\Omega^2}\left( \delta_{ab} + \alpha \times \frac{3}{2}  M_{\rm P}^2 
  \frac{\partial_a \Omega^2 \partial_b \Omega^2}{\Omega^2}\right)
\ee 
denotes a field-space metric with $\alpha=1$ for the metric case and $\alpha=0$ for the Palatini case. The connection $\Gamma_E$ can again be identified with the Levi-Civita connection $\bar \Gamma$.

A simple computation of the Gauss curvature associated  to the field manifold $\gamma_{ab}$ reveals that this is generically different from zero, forbidding the reduction of Eq.~\eqref{einsteinframe} to a fully \textit{canonical} form. The kinetic sector in Eq.~\eqref{einsteinframe} can be, however, diagonalized if the relation among the non-minimal gravitational coupling and the usual Einstein-Hilbert term is highly hierarchical. In particular, if $ \xi_{s} s^2+\xi_h h^2 \ll M_\text{P}^2$, the Weyl factor in Eq.~\eqref{Omega1}  equals one and there is essentially no difference among the two Weyl-related frames.  On the contrary, if $ \xi_{s} s^2+\xi_h h^2 \gg M_\text{P}^2$ the model coincides in form with the Higgs-Dilaton model \cite{GarciaBellido:2011de} and can be diagonalized by the field redefinition in Ref.~\cite{Casas:2017wjh}, namely\footnote{Note that the \textit{angular} variable $\Theta$ in Eq.~\eqref{eq:Thetadef} is invariant under the simultaneous rescaling of $s$ and $h$.}
\begin{eqnarray}
\gamma^{-2} \Theta &\equiv& \frac{(1+6\,\alpha \, \xi_{s})s^2+(1+6 \,\alpha \, \xi_{h})h^2}{\xi_{s} s^2+\xi_{h} h^2}\,,\label{eq:Thetadef} \\
\exp\left[\frac{2\gamma\Phi}{M_{\rm P}}\right] &\equiv& \frac{\kappa_c}{\kappa} \frac{(1+6 \,\alpha \,  \xi_{s})s^2+(1+6 \,\alpha \, \xi_{h}) h^2}{M_{\rm P}^2}\,,\label{eq:Phidef} 
\end{eqnarray}
with
\be\label{kappadef}
\kappa_c \equiv-\frac{\xi_{s}}{1+6\,\alpha\,\xi_{s}}\,, \hspace{10mm} 
\kappa \equiv\kappa_c\left(1-\frac{\xi_{h}}{\xi_{s}}\right)\,, \hspace{10mm} \gamma \equiv \sqrt{\frac{\xi_{h}}{1+6\,\alpha\,\xi_{h}}}\,.
\ee
In terms of the new coordinates, the Einstein-frame action takes the very simple form
\begin{equation}\label{action_HD1}
	S_E = \int d^4x \sqrt{-g}\left[  \frac{M_{\rm P}^2}{2}R 
-\frac{K(\Theta)}{2}(\partial\Theta)^{2} 
- \frac{G(\Theta)}{2}(\partial \Phi)^2 
 -U(\Theta)\right]\,,
\end{equation}
with 
\begin{eqnarray}\label{poles} 
U(\Theta)&=& U_0(1-\Theta)^2 \left[1+\frac{\lambda_{hs}}{\lambda} X(\Theta)+\frac{\lambda_h}{\lambda}X^2(\Theta)\right]\,,   \\
X(\Theta)&=&\frac{\kappa_c}{\kappa-\kappa_c}\frac{\Theta-\sigma}{\Theta-1}\,,\\
K(\Theta)&=&-\frac{M_{\rm P}^2}{4\, \Theta}\left(\frac{1}{\kappa\Theta+c}+\frac{a}{1-\Theta}\right)\,, \label{eq:K}\\
G(\Theta)&=&\Theta, \label{eq:G}
\end{eqnarray}
and
\be\label{Utheta}
U_0\equiv\frac{\lambda_{s} \, a^2 M_{\rm P}^4}{4}\,, \,\hspace{10mm} \sigma=\frac{1}{a\,\kappa}\frac{\kappa_c-\kappa}{\kappa_c}\,, \hspace{10mm}
a \equiv \frac{1+6 \,\alpha\,\kappa}{\kappa}\,,\hspace{10mm} c\equiv \frac{\kappa}{\kappa_c}\gamma^2\,,
\ee
The structure of Eq.~\eqref{action_HD1} is particularly enlightening:
\begin{enumerate}
\item The functions $U$, $K$ and $G$ are  $\Phi$-independent. The emergent shift symmetry $\Phi\to \Phi+C$ can be understood as the non-linear realization of the approximate dilatation symmetry of Eq.~\eqref{nonminimal_action} at $ \xi_{s} s^2+\xi_h h^2 \gg M_\text{P}^2$, with $\Phi$ the associated Goldstone boson or \textit{dilaton}.  In the presence of this symmetry, the conservation of the dilatation current
\begin{equation}\label{eq:conserv}
\frac{1}{a^3}\frac{d}{d t}\left(a^3 \gamma_{ab}\dot\varphi^a\Delta\varphi^b\right)\simeq 0\,,
\end{equation}
-- with the dots standing for derivatives with respect to the coordinate time $t$ and $\Delta\varphi^a$ denoting the infinitesimal action of dilatations on the fields -- leads to a rapid freezing of the dilaton field $\Phi$~\cite{GarciaBellido:2011de,Casas:2018fum}.  This emergent property forces both $h$ and $s$ to participate in inflation, describing an elliptical trajectory in the $\lbrace h,s \rbrace $ plane.
Note, however, that, the inflationary dynamics is essentially governed by a single variable $\Theta$, thus preventing the generation of large isocurvature fluctuations and/or non-Gaussianities~\cite{GarciaBellido:2011de,Casas:2018fum}. From this point of view, the numerical results of Ref.~\cite{Carrilho:2018ffi} are not surprising, but rather a natural consequence of the emergent scale symmetry in the large field regime. 

\item  The function $K(\Theta)$ contains three poles at $\Theta=0$, $\Theta=-c/\kappa$ and $\Theta=1$. The last one is a `Minkowski' pole around which the usual SM minimally coupled to gravity is approximately recovered~\cite{Karananas:2016kyt,Casas:2018fum}.
The poles at  $\Theta=0$ and $\Theta=-c/\kappa$ lead to an effective stretching of the canonically normalized variable 
\begin{equation}
\theta=\int^\theta \frac{d\Theta}{\sqrt{-4\,  \Theta
(\kappa \Theta+c)}}\,,
\end{equation}
which allows for inflation even if the potential $U(\Theta)$ is not sufficiently flat~\cite{Karananas:2016kyt,Casas:2018fum}. For $\xi_h=0$ ($c=0$), the pole at $\Theta=0$ is quadratic and one recovers the standard exponential stretching appearing in single-field DM-driven inflation~\cite{Liddle:2006qz,Cardenas:2007xh,Panotopoulos:2007ri,Liddle:2008bm,Lerner:2009xg,Bose,DeSantiago:2011qb,Khoze:2013uia,Bastero-Gil:2015lga,Kahlhoefer:2015jma,Tenkanen:2016twd,Choubey:2017hsq}, 
\begin{equation}
\Theta =
\exp\left(-2\sqrt{-\kappa}\,\theta\right)\,. 
\end{equation}
For non-vanishing values of $\xi_h$ ($c\neq 0$), the inflationary pole at $\Theta=0$ is no longer reachable and we are left with a linear pole at $\Theta=-c/\kappa$. In this case, the stretching of $\theta$ is restricted to a compact field range,
\begin{equation}
\Theta =
\frac{c}{-\kappa}\cosh (\sqrt{-\kappa}\,\theta) \,.
\end{equation}
\end{enumerate}
For the field values relevant for inflation, the `Minkowski' pole at $\Theta=1$ in Eq.~\eqref{eq:K} can be safely neglected.  In this limit, we are left with a very simple action,
\begin{equation}\label{action_HD2}
	S_E = \int d^4x \sqrt{-g}\left[  \frac{M_{\rm P}^2}{2}R 
-\frac{1}{2}\left(-\frac{M_{\rm P}^2  (\partial\Theta)^{2}}{4\, \Theta (\kappa\Theta+c)}
+ \Theta (\partial \Phi)^2\right) -U(\Theta)\right]\,,
\end{equation}
displaying a maximally symmetric field-derivative manifold with Gaussian curvature $\kappa$~\cite{Casas:2018fum,Karananas:2016kyt}.  The highly symmetric structure of  Eq.~\eqref{action_HD2}  has a strong impact on the inflationary observables 
\begin{eqnarray} \label{asrdef}
n_s=1+ 2\eta-6\epsilon\,, \hspace{10mm} \alpha_s=8\epsilon(2\eta-3\epsilon)-2\delta, \hspace{17mm} r= 16\epsilon \,,
\end{eqnarray}
with $n_s$ and $\alpha_s$ the spectral tilt and running  of the primordial spectrum of scalar perturbations,  respectively, $r$ the tensor-to-scalar ratio and $\epsilon$, $\eta$ and $\delta$ the slow-roll parameters (see below). In particular, the exponential stretching of the canonically normalized variable in the vicinity of the kinetic poles makes these quantities almost insensitive to the precise potential shape, provided that this is analytic around $\Theta=0$ and $\Theta=-c/\kappa$. The global amplitude of the potential remains, however, a free parameter to be fixed by the amplitude of primordial scalar perturbations
\begin{equation}
A_s=\frac{1}{24\pi^2 M_{\rm P}^4}\frac{U}{\epsilon}\,.
\end{equation}
As discussed in Sec.~\ref{inflation}, we are mainly interested in an scenario in which the scalar potential~\eqref{potential} is dominated by the $\lambda_s s^4/4$ term. As emphasized in the main text, this can be achieved either if $h=0$ or if the last two terms in Eq.~\eqref{poles} are small in the vicinity of the $\Theta=0$ pole. The second possibility corresponds to a parameter choice
\begin{equation}
\lambda_s\gg  \frac{\vert \lambda_{hs}\vert }{1+6\alpha\kappa}\,,\, \frac{\lambda_h}{(1+6\alpha \kappa)^2}\,,
\end{equation}
which in the limit $\xi_s\gg \xi_h$ gives the conditions
\begin{equation}
\label{lambda_hierarchy}
\lambda_s\gg \vert \lambda_{hs}\vert\, (1+6\alpha \xi_s),\,\lambda_h\,(1+6\alpha\xi_s)^2\,.
\end{equation}
As discussed in the main text, this hierarchy is not compatible with the requirement that the $s$ field acts as {\it both} the inflaton and a stable FIMP particle constituting all observed DM, unless $\lambda_h \ll \lambda_s \simeq 10^{-8}$ at the inflationary scale, which we consider to be a highly fine-tuned scenario but nevertheless present it here for completeness. From the inflationary side itself, however, there is no problem with having the hierarchy \eqref{lambda_hierarchy}.

For $h=0$ we retain the results presented in Sec.~\ref{inflation}. For $h\neq 0$ and the above hierarchy among couplings, the inflationary observables following from \eqref{action_HD2} with the approximate potential 
\begin{equation}\label{Uthetaapprox}
U(\Theta)\simeq U_0(1-\Theta)^2 \,,
\end{equation}
can be computed either in terms of the usual slow-roll parameters for the canonically normalized field $\chi \equiv \int \sqrt{K(\Theta)} d\Theta$ [cf. \eqref{SRparameters1}] or using  the modified slow-roll parameters
\be\label{epsdef}
\epsilon\equiv\frac{M_{\rm P}^2}{2 K}\left(\frac{U_{,\Theta}}{U}\right)^2\,, \hspace{7mm}
\eta \equiv \frac{M_{\rm P}^2}{\sqrt{K}U}\left(\frac{U_{,\Theta}}{\sqrt{K}}\right)_{,\Theta} \,, \hspace{7mm}
\delta\equiv\frac{M_{\rm P}^4 U_{,{\Theta}}}{ K U^2}\left[\frac{1}{\sqrt{K}}
\left(\frac{U_{,\Theta}}{\sqrt{K}}\right)_{,\Theta}\right]_{,\Theta}\,,
\ee
for the non-canonical inflaton field $\Theta$. We follow here the second approach.  The inflationary observables \eqref{asrdef} 
are understood to be evaluated at a field value $\Theta_*\equiv \Theta(N_*)$ with 
\begin{equation}\label{Nefolds}
N_*=\frac{1}{M_\text{P}}\int_{\Theta_{\rm E}}^{\Theta_*}\frac{\sqrt{K}d \Theta}{\sqrt{2\epsilon}}=\frac{1}{8c}\ln\left[\frac{\Theta_*}{\Theta_{\rm E}}\left(\frac{\kappa \Theta_{\rm E}+c}{\kappa\Theta_*+c}
  \right)^{1+\frac{c}{\kappa}}\right]
\end{equation}
the number of $e$-folds, and
\begin{equation}
\Theta_{\rm E}=\frac{1-4c-2\sqrt{4c^2-2c-2\kappa}}{1+8\kappa}
\end{equation}
the value of the field at the end of inflation, $\epsilon(\Theta_{\rm E})\equiv1$. Since we are especially interested in a scenario in which the Higgs field plays a subdominant role during inflation ($\xi_{h}\ll \xi_{s}$ or $h=0$), we will approximate  $\vert \kappa\vert\simeq \vert\kappa_c\vert$ and assume the ratio $c/\vert \kappa\vert$ to be small. This allows us to analytically invert  Eq.~\eqref{Nefolds} while keeping track of the leading order effects of $\xi_{h}$. To the lowest order in $c/\vert\kappa\vert$ we obtain the following analytical expressions for the amplitude of the primordial scalar perturbations~\cite{Casas:2017wjh}, 
\begin{equation}\label{As}
A_s
 =\frac{\lambda_{s} \sinh^{2} \left(4 c N_* \right)}{1152\pi^2 \xi^2_{\rm eff}\,  c^2} \,,\hspace{10mm} 
 \xi_{\rm eff}\equiv\frac{1}{\sqrt{6 a^2 \vert\kappa_c\vert}}\,,
\end{equation}
its spectral tilt and running
\begin{eqnarray}
n_s &=& 1-8\,c \coth\left(4 c N_*\right)\,,   \hspace{10mm}  \alpha_s = - 32\, c^2 \sinh^{-2} \left(4 c N_* \right)\,, \label{nsr1b} 
\end{eqnarray}
and the tensor-to-scalar ratio
\be\label{nsr2}
r=  \frac{32\, c^2}{\vert \kappa_c\vert} \sinh^{-2} \left(4 c N_*\right)\,.
\ee 
We can distinguish two asymptotic regimes. For $4 c N_*\gg 1$, the spectral tilt approaches $-\infty$  while its running and the tensor-to-scalar tend to zero, in good agreement with the single-pole behavior of Eq.~\eqref{action_HD2} at $c=0$~\cite{Terada:2016nqg}. On the other hand, for $4 c N_*\ll 1$, the above expressions converge to the 
single-field results found in previous studies~\cite{Bezrukov:2007ep,Bauer:2008zj}, namely
\be 
r\simeq \frac{2}{\vert \kappa_c\vert N_*^2}\,, \hspace{15mm}
n_s \simeq 1-\frac{2}{N_*} \,,  \hspace{15mm} 
\alpha_s \simeq  - \frac{2}{N_*^2}\,.
\ee 
and
\be
\label{xilambda_relation}
A_s
 =\frac{\lambda_{s}N_*^2(1+6\alpha\kappa_c)^2}{12\pi^2|\kappa_c|}\,,
\ee
with  $\vert\kappa_c\vert\simeq 1/6$ at small $\xi_{h}$ in the metric case ($\alpha=1$) and  $\kappa_c\simeq  \xi_{s}$ at small $\xi_{h}$ in the Palatini case ($\alpha=0$).


\section{Inflaton decay rates}
\label{decayrates_appendix}

In this appendix we discuss the effective semi-perturbative decay rates of the inflaton condensate following Refs.~\cite{Tenkanen:2016twd, Kainulainen:2016vzv} (see also Refs.~\cite{Abbott:1982hn,Ichikawa:2008ne,Nurmi:2015ema}).

Once the inflaton condensate starts to oscillate around the minimum of its quartic potential, it evolves according to
\be 
s(t)=s^{(4)}_0(t)\,{\rm cn}\left[0.85\sqrt{\lambda_s}s^{(4)}_0(t)(t-t_{\rm end}),\,1/\sqrt{2}\right] ,
\ee
with cn the elliptic cosine function, $s^{(4)}_0(t)=s_{\rm end}\sqrt{t_{\rm end}/t}$ a time-dependent oscillation amplitude, $s_{\rm end}$ the field value at the end of inflation and $t$ the cosmic time. Due to the expansion of the Universe the field relaxes to smaller field values, reaching eventually the region where the potential is approximately quadratic.  When that happens the field evolves according to
\be
s(t) = s^{(2)}_0(t)\cos\left(\mu_s t\right) ,
\ee
with $s^{(2)}_0(t)$ another time-dependent oscillation amplitude.  The oscillating background generates an additional time-dependent mass term for $s$ and $h$ particles,
\be
\label{adiabatic_masses}
\begin{aligned}
M_{s}^2 &= \mu^2_{s} + 3\lambda_{s} s(t)^2, \hspace{15mm}
M_{h}^2 &= \mu_{h}^2 + \frac{\lambda_{hs}}{2} s(t)^2 , 
\end{aligned}
\ee
which is the origin of particle production.\footnote{In all our considerations, we neglect possible thermal corrections to masses.} 
It is convenient to expand the field in terms of Fourier modes as
\be
\label{s_fourier}
s^2(t) = \sum_{n=-\infty}^{\infty}\zeta_n e^{-i2\omega nt} ,
\ee
with $\omega\simeq 0.85\sqrt{\lambda_s}s_0$ for the quartic potential and $\omega=\mu_s$ for the quadratic one. Following Refs. \cite{Ichikawa:2008ne,Nurmi:2015ema,Kainulainen:2016vzv}, we take the interaction terms in the Lagrangian to be
\be
\lambda_s s^2(t)\int {\rm d}^3x \hat{s}\hat{s}, \hspace{15mm}
\lambda_{hs} s^2(t)\int {\rm d}^3x \hat{h}\hat{h},
\ee
where $\hat{s}$, $\hat{h}$ are quantized fields with the usual commutation relations. Using these, we can compute the transition amplitude from the initial state with no particles to the final two-particle state in the usual way, as in Refs.~\cite{Ichikawa:2008ne,Nurmi:2015ema,Kainulainen:2016vzv}. The decay rates of the condensate energy density induced by the above interactions are then given by
\be
\begin{aligned}
\Gamma_{s_0\rightarrow ss} &= \frac{9\lambda_{s}^2 \omega}{8\pi\rho_{s_0}}\sum_{n=1}^{\infty}n|\zeta_n |^2\sqrt{1-\left(\frac{M_{s}}{n\omega}\right)^2} ,  \\
\Gamma_{s_0\rightarrow hh} &=\frac{\lambda_{hs}^{2}\omega}{8\pi\rho_{s_0}}\sum_{n=1}^{\infty}n|\zeta_n |^2 \sqrt{1-\left(\frac{M_h}{n\omega}\right)^2} ,
\end{aligned}
\label{eq:decayrates_appendix}
\ee
where $\rho_{s_0}$ is the average energy density of the field in the corresponding potential region, namely 
$\rho_{s_0}=\lambda_s s_0^4/4$ for the quartic potential and $\rho_{s_0}=\mu_s^2s_0^2/2$ for the quadratic one.\footnote{In order to simplify the notation, we refer to the oscillation amplitudes $s^{(4)}_0(t)$ and $s^{(2)}_0(t)$ as $s_0$.} Finally, we average the decay rates over one oscillation cycle, which gives
\be
\Gamma^{(4)}_{s_0\to ss} = 0.023\lambda_{s}^{\frac32}s_0 , \hspace{20mm}
\Gamma^{(4)}_{s_0\to hh} = 0.002\lambda_{hs}^2\lambda_{s}^{-\frac12}s_0 ,
\label{eq:decayrates_appendix2}
\ee
in the quartic potential, and
\begin{equation}
	\Gamma^{(2)}_{s_0\to hh} = \frac{\lambda_{hs}^2s_0^2}{64\pi m_{s}}\sqrt{1-\left(\frac{m_{h}}{m_{s}}\right)^{2}} ,
\end{equation}
in the quadratic one. The decay rates~\eqref{eq:decayrates_appendix2} neglect the bare mass terms but account for the adiabatic mass contributions~\eqref{adiabatic_masses}, and are thus expected to describe the dynamics of reheating to a sufficient accuracy. In evaluating $\Gamma^{(4)}_{s_0\to hh}$, we have assumed a hierarchy of couplings $\lambda_{hs}=10\lambda_s$, which affects the result via the suppression that the effective Higgs mass $M_h$ generates for Eq.~\eqref{eq:decayrates_appendix}. Because the correct treatment of the time-dependent effective masses is a non-trivial open issue and the main origin of discrepancy between different results in the literature (see e.g. Refs.~\cite{Ichikawa:2008ne, Cosme:2017cxk}, where the effective masses where neglected, and Ref.~\cite{Nurmi:2015ema} where the mass terms we treated differently from the Ref.~\cite{Kainulainen:2016vzv} followed here), the uncertainty related to the chosen hierarchy of couplings is not expected to pose a major issue; for instance, the choice $\lambda_{hs}=10^3\lambda_s$ would give the prefactor $0.0002$ instead of $0.002$. This discrepancy is smaller than what different choices of effective masses generically cause. Resolving the issue with such mass terms is, however, beyond this study. 

\section{Bounds on $\lambda_s$ from reheating}
\label{appendix}

In this appendix we calculate bounds on $\lambda_{s}$ from the reheating stage assuming that most of DM is produced by freeze-in at a later stage in cosmic history, i.e. after reheating. The bounds stem from requiring that while the inflaton condensate is oscillating, it transfers most of its energy into the SM sector and only a small amount of its energy is distributed to finite-momentum quanta of the $s$ field. In the following, we will compute two different bounds, one for the case where reheating occurs while the inflaton is still oscillating in the quartic part of its effective potential, and one for the case where reheating occurs once the field has relaxed to the quadratic part.

\subsection*{The quartic case}

Let us first assume that reheating occurs in the quartic regime. We can then estimate that at the time of reheating the energy density in $s$ particles is
\be
\rho_{s}(T_{\rm RH}) \simeq {\rm BR}\,\rho_{s_0}(T_{\rm RH}) ,
\ee
where ${\rm BR}= 11.5\lambda_{s}^2/\lambda_{hs}^2$ is the inflaton branching ratio into two $s$ particles
\be
\label{br}
{\rm BR} =  \frac{\Gamma_{s_0\to ss}}{\Gamma_{s_0\to ss}+\Gamma_{s_0\to hh}}\simeq \frac{\Gamma_{s_0\to ss}}{\Gamma_{s_0\to hh}}\,,
\ee
and $\rho_{s_0}(T_{\rm RH})$ is the total energy density of the inflaton condensate at the time of reheating. Because in our case the inflaton  decay produces $s$ particles with an almost monochromatic spectrum with momentum $k=\sqrt{3\lambda_{s}}s_0$ and the particles only redshift after their production~\cite{Kainulainen:2016vzv}, we can estimate that they become non-relativistic at
\be
a_{\rm nrel} = \frac{\sqrt{3\lambda_{s}}s_{\rm end}}{m_{s}} = \left(108\lambda_{s}\right)^{\frac14}\frac{\sqrt{H_{\rm end}M_{\rm P}}}{m_{s}} ,
\ee
where the subscript `end' refers to the value of each quantity at the end of inflation. After the $s$ particles have become non-relativistic, their energy density scales as
\be
\rho_{s}(a) = {\rm BR}\,\rho_{s_0}(a_{\rm end})\left(\frac{a_{\rm end}}{a_{\rm nrel}}\right)^4\left(\frac{a_{\rm nrel}}{a}\right)^3 
= \left(\frac{3}{4\lambda_{s}}\right)^{\frac14}{\rm BR}\,\frac{(H_{\rm end}M_{\rm P})^{\frac32}m_{s}}{a^3} .
\ee
Here we have normalized the scale factor such that $a_{\rm end}=1$. By demanding that at the time of matter-radiation equality the $s$ particles only constitute less than a fraction $C$ of the total DM abundance, i.e.
\be
\label{sparticle_criterion}
\rho_{s}(a_{\rm eq})  \leq C\times \frac{3H_{\rm eq}^2M_{\rm P}^2}{2\left(1+\frac{\Omega_{\rm b}}{\Omega_{\rm DM}}\right)} ,
\ee
with $a^2_{\rm eq}=g_*^{\frac12}(a_{\rm eq})g_*^{\frac16}(a_{\rm end})g_{*S}^{-\frac23}(a_{\rm eq})H_{\rm end}/H_{\rm eq}$ ,
we obtain
\be
\label{lambdaCondition}
\lambda_{s} \leq  6\times 10^{-7} \lambda_{hs}^{\frac87}\left(\frac{{\rm GeV}}{m_{s}}\right)^{4/7}\left(\frac{C}{0.1}\right)^{4/7} ,
\ee
which is the result presented in Eq.~\eqref{hierarchy}. To obtain a numerical value for the prefactor, we used $H_{\rm eq}=1.5\times 10^{-37}$ GeV, $\Omega_{\rm b}=0.02$, $\Omega_{\rm DM}=0.12$, $g_*(a_{\rm eq})=3.363$, $g_{*S}(a_{\rm eq})=3.909$, and $g_*(a_{\rm end})=106.75$~\cite{Aghanim:2018eyx}.

\subsection*{The quadratic case}

Let us then assume that reheating occurs in the quadratic regime. In that case, the maximum energy density in $s$ particles produced while the inflaton was still oscillating in the quartic regime\footnote{Recall that in the quadratic regime the channel $s_0\to ss$ is kinematically blocked.} is
\be
\rho_{s} = \left.\frac{\Gamma^{(4)}_{s_0\to ss}\rho_{s_0}}{H}\right|_{a_{\rm trans}} = 0.004\sqrt{\lambda_{s}}M_{\rm P}m_{s}^3 ,
\ee
because $\Gamma^{(4)}_{s_0\to ss}/H$ reaches its maximum at the transition value $a_{\rm trans}$ where the two potential terms become equal, $\lambda_{s}s_0^2(a_{\rm trans}) = 2m^2_{s}$. Since after this point the particles are non-relativistic, we obtain
\be
\rho_{s}(a) = \left. \frac{\Gamma^{(4)}_{s_0\to ss}\rho_{s_0}}{H}\right|_{a_{\rm trans}} \left(\frac{a_{\rm trans}}{a}\right)^3 
= \left. \frac{\Gamma^{(4)}_{s_0\to ss}\rho_{s_0}}{H}\right|_{a_{\rm trans}}\frac{3\lambda_{s}M_{\rm P}^2H^2}{m_{s}^4} ,
\ee
where we used the fact that after the transition the Universe is effectively matter-dominated until reheating. By again requiring that at the time of matter-radiation equality the $s$ particles constitute less than a fraction $C$ of the total DM abundance (cf. Eq.~\eqref{sparticle_criterion}), we obtain
\be
\label{lambdaCondition2}
\lambda_{s} < 2\times 10^{-12}\left(\frac{m_{s}}{\rm GeV}\right)^{\frac23} \left(\frac{C}{0.1}\right)^{\frac23},
\ee
which is the result presented in Eq.~\eqref{hierarchy2}.

\bibliographystyle{JHEP}
\bibliography{DM_inflation}


\end{document}